\documentclass[preprint]{aastex}

\usepackage{amsmath,bm,mathrsfs}

\DeclareMathAlphabet{\tens}{OT1}{cmss}{bx}{n}

\begin{document}

\arraycolsep .15em

\title{Analysis of Seeing-Induced Polarization Cross-Talk and\break
Modulation Scheme Performance}

\author{R.\ Casini, A.\ G.\ de Wijn, P.\ G.\ Judge}

\affil{High Altitude Observatory,
National Center for Atmospheric Research,\footnote{The National Center
for Atmospheric Research is sponsored by the National Science
Foundation.}\break
P.\ O.\ Box 3000, Boulder, CO 80307-3000, U.S.A.}

\begin{abstract}
      We analyze the generation of polarization cross-talk in Stokes 
polarimeters by atmospheric 
seeing, and its effects on the noise statistics of spectro-polarimetric 
measurements for both single-beam and dual-beam instruments.
      We investigate the time evolution of seeing-induced correlations 
between different states of one modulation cycle, and compare the response 
to these correlations of two popular polarization modulation schemes in
a dual-beam system.
      Extension of the formalism to encompass an arbitrary number of
modulation cycles enables us to compare our results with earlier work.
      Even though we discuss examples pertinent to solar physics,
the general treatment of the subject and its fundamental results might
be useful to a wider community.
\end{abstract}

\section{Introduction}

      Many important solar phenomena are driven by the interaction 
of turbulent plasma with magnetic fields.
      Measurements of the full state of polarization of solar spectral 
lines are routinely used to infer properties of solar vector magnetic 
fields from their imprints in the emergent spectra through the Zeeman 
and Hanle effects.
      All ground-based measurements are detrimentally affected by image 
distortion introduced by atmospheric seeing, which has significant power 
at frequencies up to at least $100\,$Hz, significantly higher than the 
frame rates of typical CCD cameras.
      The seeing-induced errors due to image distortion during 
an observation are expected to increase as the telescope aperture 
increases in size, since the telescope can then resolve finer 
structures and hence observe steeper gradients in images. 

      The Advanced Technology Solar Telescope \cite[ATST;][]{Ri08}, 
a 4-m aperture, off-axis telescope, will soon enter construction 
phase. The European Solar Telescope \cite[EST;][]{Co08} is also a 
4-m class telescope currently under design.
      Spectro-polarimetry is the prime mode of operation of this new 
generation of solar telescopes, but there is an urgent need to identify 
potential pitfalls before committing to a particular design for the 
modulation and detection of polarized light using such telescopes.
      Space-based spectro-polarimetric instruments, like those on-board the 
Hinode \cite[Solar-B;][]{Ts08} and the Solar Dynamics Observatory
\cite[SDO;][]{Sc12} spacecrafts, are by necessity fed by smaller telescopes, 
and while their observations profit greatly from the absence of 
atmospheric disturbances, they are still affected by residual image 
motion due to spacecraft jitter.

      The purpose of the present paper is to re-examine and extend 
earlier theoretical studies of polarization errors introduced 
by the effects of atmospheric seeing.
Potentially, this can help refine the instrument requirements for 
the polarization systems of these large, multi-national projects, as well 
as of future solar space missions,
such as Solar-C \citep{Sh11}.

      Two modes of operation are of particular interest for our study.
      Slit-based spectro-polarimeters often use long integration times, 
including a number of modulation cycles that is typically of the order of 
a few tens.
      Because of the need to scan across the spectral line domain,
      tunable imaging spectro-polarimeters typically implement 
      much fewer measurements of the modulated intensities -- often 
      just one modulation cycle -- for each wavelength position.
      The formalism developed in this paper is general, and can be applied 
to both types of instruments and observations.
A substantial difference between the two types of observations
is that post-processing techniques such as image shifting and de-stretching 
can be effectively adopted to partially correct for seeing-induced 
distortion in observations done with imaging polarimeters, whereas this 
option is not available in the case of slit-based spectro-polarimeters.

      A separate question -- also of fundamental importance for
observational spectro-polarimetery -- concerns the effects of
polarization calibration errors on the degree of polarization cross-talk
affecting an observation. This subject has been exhaustively treated by
\cite{AC08}, and therefore it will not be addressed here.

      The plan of this paper is the following.
      In Sect.~\ref{sec:seeing}, we describe a model for the
effects of atmospheric seeing on the polarization signals detected by an
observer. 
	In Sect.~\ref{sec:formalism}, we lay the foundation of the formalism.  
	Using the point of view of the statistics of random processes we 
define the fundamental statistical observables for spectro-polarimetry
in the presence of atmospheric seeing.
      In Sect.~\ref{sec:dualbeam}, this formalism is extended to the case 
of dual-beam polarimetry, emphasizing those aspects of the problem  where 
this extension is not trivial.
      In Sect.~\ref{sec:covariances} we make use of the conceptual 
framework of the statistics of stationary random processes to gain
a deeper understanding of
the effects of seeing correlations on spectro-polarimetric measurements.
      In particular, in that section we study qualitatively the effect 
of adaptive optics (AO) corrections on the temporal decay of seeing 
correlations.
      Those results are successively applied to study the performance of 
two popular modulation schemes in the presence of seeing.
      In Sect.~\ref{sec:generalization}, these results are further 
generalized to the case in which seeing correlations extend over the 
duration of one elemental observation (e.g., one slit position), and the 
effects of seeing on the performance of optimally efficient modulation 
schemes are studied both in the absence and in the presence of low-order 
(i.e., tip-tilt) AO corrections (based on observed data).
      Finally, we compare our results and conclusions with those from 
previous works on the subject.

\section{Polarization and atmospheric seeing}
\label{sec:seeing}

      We conventionally describe the polarization state of a radiation 
beam by its Stokes vector, $\bm{S}\equiv(S_1,S_2,S_3,S_4)$, where $S_1$ 
is the intensity of radiation, $S_2$ and $S_3$ describe the two independent 
states of linear polarization on the plane normal to the propagation
direction, and finally $S_4$ is the parameter for circular polarization.

      Because photon detectors are practically sensitive only to the 
intensity of the incoming radiation, the measurement of polarized radiation 
requires that its states of polarization be encoded into intensity signals.
      This encoding process is conventionally called \textit{modulation}, 
and it is most often achieved through time-varying optical devices 
(modulators) that modify in a known way the polarization states of the 
incoming radiation.

      Let $\bm{m}(t)$ be the four-vector describing the polarization modulation 
process for the four Stokes parameters,
so that $\bm{m}(t)\cdot\bm{S}$ gives the detected signal of the
modulated intensity at time $t$ (see Eq.~[\ref{eq:signal_i}] below).
      We will assume that this modulation vector is perfectly known (either 
by design or calibration), and that it is a periodic function of time, so 
that $\bm{m}(t+\tau)=\bm{m}(t)$, where $\tau$ is the period of the 
modulation cycle.
      Because of these properties, the time process of polarization 
modulation is both \emph{deterministic} and \emph{stationary}
in a statistical sense.
      Evidently one needs at least four independent measurements 
during a modulation cycle to fully determine the Stokes vector. 
      The actual number of measurements $n$ taken during a given modulation 
cycle represents the number of modulation states for that cycle.

      We also assume that the physical conditions of the light emitting 
region are stationary during the time interval $T$ needed to perform 
a \emph{complete} spectro-polarimetric observation of a given element 
of spatial or spectral sampling, with the required sensitivity.
	In what follows, the term ``complete observation'' will always refer 
to such elemental operation, for instance, the integration of the light
signal at one slit position of a spatial scan, in the case of a 
grating-based spectro-polarimeter, or at one wavelength position of a 
spectral scan, in the case of a tunable, imaging spectro-polarimeter. 
	Evidently, such observation must always include at least one full 
modulation cycle.

      In the presence of atmospheric seeing, a given pixel on the detector 
collects photons from different areas of the observed region, with a 
characteristic time scale $t_0$ ($\sim 0.01\,$s).
	The cause for this ``smearing'' of the image can be a true 
displacement of the target by the seeing during the integration time, 
or seeing-induced distortions of the effective PSF at the detector,
which determine a time-varying weighing of different elements of the 
observed region. This time-dependent smearing is in addition to the 
one determined by diffraction because of the finite aperture of the 
telescope. 
As a result, the combined effect of the atmosphere and the telescope 
is to smoothen the Stokes gradients present in the observed region, 
through a smearing length that is determined by the spatial resolution of 
the observation.

	In order to demonstrate this intuitive result, we begin by 
considering the ideal case of a perfectly stigmatic optical system 
\citep[e.g.,][]{BW65} observing in the absence of atmospheric seeing. 
	In this case, a 1-1 mapping can be established between the object and 
the image planes. 
	We indicate with $\bm{\tilde S}_{ij}(t)$ the Stokes vector
falling on the \emph{resolution element}\footnote{%
Typically, the pixel size in an optical system is matched to the 
critical sampling width (both spatial and spectral) determined in
accordance to the Nyquist criterion \cite[e.g.,][]{Go96}, assuming
diffraction-limit performance of the instrument. Thus, the
resolution element will span at least $2\times2$ pixels, or a
larger number for performance worse than the diffraction limit.}
of coordinates $(i,j)$ on the image plane, at a given time $t$.
	Similarly, we indicate with $\bm{S}_{ij}$ the Stokes vector emitted 
by the element on the object plane (i.e., the solar surface) that 
corresponds to the element $(i,j)$ on the detector, 
through the inverse mapping from the image plane onto the object plane. 
	Note that the Stokes vector at the source is not given as a 
function of time, in agreement with the condition stated earlier that
the time interval $T$ to perform an elemental observation should be much 
shorter than the typical evolution time of the observed solar structure.

	As customary, we assume that the transport of radiation from the 
object plane to the image plane is described by a linear operator.
Hence, we can write
\begin{equation} \label{eq:model0}
\bm{\tilde S}_{ij}(t)=\sum_{kl}\tens{T}_{ij}^{kl}(t)\,\bm{S}_{kl}\;,
\end{equation}
where $\tens{T}_{ij}^{kl}(t)$ is the transfer (Mueller) matrix at time
$t$ of the imaging system telescope+atmosphere, which maps the element 
$(k,l)$ on the object plane to the corresponding resolution element 
$(i,j)$ on the image plane. 
	In the ideal case of a perfectly stigmatic telescope, and in the
absence of atmospheric seeing, evidently
\begin{equation} \label{eq:model.diag}
\tens{T}_{ij}^{kl}(t)=\delta_i^k\delta_j^l\,\tens{T}_{ij}(t)\;,
\end{equation}
where $\tens{T}_{ij}(t)$ is the transfer matrix of the imaging system,
pertaining to the element $(i,j)$ of the field of view.

	The general form of Eq.~(\ref{eq:model0}) applies in the
presence of smearing of the image, which may be caused by atmospheric 
seeing, by the instrument's PSF, as well as by the optical aberrations 
of the imaging system.\footnote{%
Of course, the discretization of the object plane implied by 
Eq.~(\ref{eq:model0}) can only be an approximation, under these more
general observing conditions.}
	The Stokes vector measured by the detector is thus properly
expressed as a function of the time, because of the variability of 
the atmosphere. 
	Possible temporal variations of the instrument -- for 
example, due to the changing optical configuration of the telescope 
during tracking of the target -- are not a concern here, since their
time scale is typically much larger than $T$. Thus, in practice, the
transfer matrix on the RHS of Eq.~(\ref{eq:model.diag}) can be assumed
to be constant.

	When observations are affected by atmospheric seeing, 
the spatial resolution -- and thus the size of the resolution element 
on the detector -- depends critically 
on the particular observing conditions, such as the possible presence of 
AO corrections, and the temporal resolution of the observation (i.e., 
the exposure time), because of the effects of image smearing mentioned 
above. 
	For example, in the limit of long time exposures, and
in the absence of AO corrections, the spatial resolution is completely
determined by the Fried parameter $r_0$ of the atmospheric seeing
\citep{Fr65}.
	According to the Van Cittert-Zernike theorem
(\citealt{VC34,Ze38}; see also \citealt{MW95}), distinct resolution 
elements can be regarded as incoherent light sources.
	The linear superposition of the source Stokes vectors expressed 
by Eq.~(\ref{eq:model0}) is thus justified by this assumption of spatial 
and temporal incoherence of the radiation emitted by resolved structures
on the object plane.

Let us now consider a point $(p,q)$ on the image plane, located inside 
a sufficiently small neighborhood of $(i,j)$,
such that we can approximate
\begin{equation}	\label{eq:model1}
\bm{\tilde S}_{pq}(t)\approx \bm{\tilde S}_{ij}(t)
	+(\bm{x}_{pq}-\bm{x}_{ij})\cdot
	\bm{\nabla\tilde S}_{ij}(t)\;.
\end{equation}
We recall that, for an arbitrary vector $\bm{v}$,
\begin{eqnarray*}
\bm{\nabla v}_{ij}
&=& (\partial_x\bm{v}_{ij})\bm{\hat e}_x
	+ (\partial_y\bm{v}_{ij})\bm{\hat e}_y \\
&\equiv& l_x^{-1}(\bm{v}_{ij}-\bm{v}_{i-1,j})\bm{\hat e}_x
	+ l_y^{-1}(\bm{v}_{ij}-\bm{v}_{i,j-1})\bm{\hat e}_y\;,
\end{eqnarray*}
where 
$l_x\equiv|\bm{x}_{ij}-\bm{x}_{i-1,j}|$ and 
$l_y\equiv|\bm{x}_{ij}-\bm{x}_{i,j-1}|$. 
We then find, from Eq.~(\ref{eq:model0}),
\begin{subequations} \label{eq:part1.0}
\begin{eqnarray}
\partial_x\bm{\tilde S}_{ij}(t)
&=& l_x^{-1}\sum_{kl}\left[ 
	\tens{T}_{ij}^{kl}(t) - \tens{T}_{i-1,j}^{kl}(t)
	\right]\bm{S}_{kl}\;, \\
\partial_y\bm{\tilde S}_{ij}(t)
&=& l_y^{-1}\sum_{kl}\left[ 
	\tens{T}_{ij}^{kl}(t) - \tens{T}_{i,j-1}^{kl}(t)
	\right]\bm{S}_{kl}\;.
\end{eqnarray}
\end{subequations}

We now assume that the transfer matrix is \emph{shift-invariant} over some 
neighborhood of the observed point.
	By definition, this condition is automatically satisfied when the 
neighborhood lies within an isoplanatic patch of the field of view 
\citep[e.g.,][]{Go96}.
	Then, Eqs.~(\ref{eq:part1.0}) can be rewritten in the following
form,
\begin{subequations} \label{eq:part1}
\begin{eqnarray}
\partial_x\bm{\tilde S}_{ij}(t)
&=& -l_x^{-1}\sum_{kl}\left[ 
	\tens{T}_{ij}^{k+1,l}(t) - \tens{T}_{ij}^{kl}(t)
	\right]\bm{S}_{kl}\;, \\
\partial_y\bm{\tilde S}_{ij}(t)
&=& -l_y^{-1}\sum_{kl}\left[ 
	\tens{T}_{ij}^{k,l+1}(t) - \tens{T}_{ij}^{kl}(t)
	\right]\bm{S}_{kl}\;.
\end{eqnarray}
\end{subequations}

Next we note that the smearing of the image at a given point is always 
bounded in practice, in the sense that, for every point $(i,j)$ on the 
image plane, there is always a positive integer $M(i,j)$ such that
%
\begin{equation} \label{eq:boundary}
\tens{T}_{ij}^{i\pm m,j\pm n}(t)\equiv 0\;,\qquad
	\textrm{if}\;\; m,n\ge M(i,j)\;.
\end{equation}

Under typical observing conditions, the isoplanatic patch is always 
larger than the contribution region to the element $(i,j)$ due to 
smearing, which is bounded according to Eq.~(\ref{eq:boundary}).
	For example, in the absence of AO corrections and for long time
exposures, the smearing length is approximately given by the spatial
resolution corresponding to $r_0$, but the region of isoplanatism
typically includes several of these resolution elements.
	Therefore, we can assume that the set of points $(k,l)$ 
contributing to Eqs.~(\ref{eq:part1}) always lies inside a region 
where the assumption of shift-invariance of the transfer matrix is 
valid. 
	This allows us to perform a summation by parts of 
Eqs.~(\ref{eq:part1}), neglecting the associated boundary terms because
of Eq.~(\ref{eq:boundary}), to find
\begin{subequations} \label{eq:part2}
\begin{eqnarray}
\partial_x\bm{\tilde S}_{ij}(t)
&=& l_x^{-1}\sum_{kl}
	\tens{T}_{ij}^{k+1,l}(t)
	\left(\bm{S}_{k+1,l}-\bm{S}_{kl}\right)
	\equiv 
	\sum_{kl} \tens{T}_{ij}^{kl}(t)\,
	\partial_x\bm{S}_{kl}\;, \\
\partial_y\bm{\tilde S}_{ij}(t)
&=& l_y^{-1}\sum_{kl}
	\tens{T}_{ij}^{k,l+1}(t) 
	\left(\bm{S}_{k,l+1}-\bm{S}_{kl}\right)
	\equiv 
	\sum_{kl} \tens{T}_{ij}^{kl}(t)\,
	\partial_y\bm{S}_{kl}\;.
\end{eqnarray}
\end{subequations}

Equations~(\ref{eq:part2}) demonstrate the anticipated result that 
the Stokes gradients present at the object plane are smeared by
the combined effect of the atmosphere and the telescope.
The characteristic length for the variation of the Stokes 
vector on the image plane is determined by the convolution of the 
physical spatial scale of the observed structure, characterized by 
$\bm{\nabla S}_{kl}$, with the spatial resolution of the 
observation, corresponding to the domain of non-nullity of 
$\tens{T}_{ij}^{kl}(t)$ (cf.\ Eq.~[\ref{eq:boundary}]).
\emph{
The validity of the linear approximation expressed by
Eq.~(\ref{eq:model1}) is therefore limited to a range of distances 
$|\bm{x}_{pq}-\bm{x}_{ij}|$ corresponding to this ``smoothed'' length 
scale,} which is approximately given by 
$|\bm{\tilde S}_{ij}|/\|\bm{\nabla\tilde S}_{ij}\|$.

Equation~(\ref{eq:model1}) provides a snapshot of the spatial 
distribution of polarization signals in a region around the element
$(i,j)$ at the time $t$. Because the phenomenon of atmospheric 
seeing can be regarded as a stationary and ergodic\footnote{%
Simply stated, \emph{ergodicity} is 
the statistical property of a dynamical system that allows to use 
time averages in place of ensemble averages. In other words, it 
is characteristic of an ergodic process to attain, during its 
temporal evolution, all of its possible configurations, with a 
probability distribution identical to that of the ensemble.}
random process \citep[e.g.,][]{MW95}, we can assume that all 
possible realizations $\bm{\tilde S}_{pq}(t)$ 
of Eq.~(\ref{eq:model1}) will eventually manifest themselves at 
$\bm{x}_0\equiv\bm{x}_{ij}$ during the evolution of the atmospheric seeing. 
In other words, we can associate a stationary random process 
$\bm{x}(t)$ to the displacement vector of the image motion, which
describes the effect of atmospheric seeing to the lowest order, such 
that (cf.\ Eq.~[\ref{eq:model1}])
\begin{equation} \label{eq:model}
\bm{\tilde S}(\bm{x}_0;t) 
\approx \bm{\tilde S}(\bm{x}_0)
	+\left(\bm{x}(t)-\bm{x}_0\right) \cdot
      \bm{\nabla\tilde S}(\bm{x}_0)\;,
\end{equation}
where $\bm{\tilde S}(\bm{x}_0)$ and 
$\bm{\nabla\tilde S}(\bm{x}_0)$ must be interpreted as the
ensemble (or long-time) averages of the quantities 
$\bm{\tilde S}_{ij}(t)$ and $\bm{\nabla\tilde S}_{ij}(t)$
appearing in Eq.~(\ref{eq:model1}).
	Because of the random nature of atmospheric seeing, the 
long-time average of $\bm{x}(t)-\bm{x}_0$ is zero, and therefore 
$\bm{\tilde S}(\bm{x}_0)$ coincides with the Stokes vector that
would be measured in the absence of atmospheric seeing.

For small telescopes with apertures of order $r_0$, the detected signal
at any given time is mostly affected by low-order aberrations of the
wavefront, so that the seeing displacement vector $\bm{x}(t)$ can be 
effectively compensated just by the tip and tilt corrections.
      For large aperture telescopes, such that $(D/r_0)^2$ is of order
10 or larger, $D$ being the telescope's diameter, the distortions of the
incoming wavefront are significantly more complex, resulting in the
characteristic dynamic pattern of \textit{speckles} on the image plane. 
Nonetheless, even in this more general case, the low-order Zernike terms 
corresponding to the tip-tilt wavefront correction still account for the 
largest part of the seeing induced effects on the image \citep{No76}.
      We note that this more general case is still captured by the 
model of Eq.~(\ref{eq:model0}).
	Thus, as long as a linear approximation of 
$\bm{\tilde S}(\bm{x};t)$ around $\bm{x}_0$ is justified,
Eq.~(\ref{eq:model}) also describes the lowest order effects of 
seeing in observations performed with a large telescope.

Of course, the requirements for the validity of the linear 
approximation (\ref{eq:model}) become more stringent as the spatial 
resolution increases with the AO correction, because of the enhanced 
contrast of the image, and the expected increased complexity of the 
structure of the solar atmosphere at smaller spatial scales. 
	In fact, as it has already been pointed out, the spatial range 
of applicability of the linear model (\ref{eq:model}) decreases like 
$|\bm{\tilde S}|/\|\bm{\nabla\tilde S}\|$. 
	On the other hand, the AO corrections effectively reduce the RMS 
displacement vector that can be associated with the residual image 
motion. Therefore, because of the 
overall scaling of the problem, we can expect that the linear model of 
Eq.~(\ref{eq:model}) will remain applicable even in the presence of 
AO corrections, at least as far as the effects of residual image motion 
are concerned.

Higher-order effects may be responsible for local fluctuations of the
image intensity not associated with image motions, which would also be
a source of polarization cross-talk, and they may even occur as image 
artifacts introduced by the same AO corrections.
	A quantitative study of these effects lies beyond the scope of 
this paper, but it could perhaps be pursued through an extension of the 
formalism presented in this work.

\section{Statistical description of the effects of atmospheric seeing}
\label{sec:formalism}

      On the basis of the conclusions of the former section, in the 
following we will assume that, at any given time within the duration 
$T$ of a particular observation, the Stokes vector 
entering the polarization modulator of the telescope
is expressed by Eq.~(\ref{eq:model}) \cite[cf.][]{Li87,Ju04}.
	For notational convenience, in the following we always assume 
$\bm{x}_0=0$, and also drop the ``$\bm\tilde{\hphantom{S}}$'' from the 
quantities of Eq.~(\ref{eq:model}).

      According to this model,
      the modulated intensity recorded by an ideal detector (i.e., 
neglecting bias and read-out noise), for the $i$-th state of the 
modulation cycle, centered around the time $t_i$, and integrated over the 
exposure time $\Delta t$, is given by
\begin{equation}	\label{eq:signal_i}
\mathscr{I}_i=\kappa\sum_{j=1}^4\,
      \int\limits_{-\Delta t/2}^{+\Delta t/2}
      m_j(t+t_i)\,[S_j+\bm{\nabla} S_j\cdot\bm{x}(t+t_i)]\;\textrm{d}t
      +\delta\mathscr{I}_i\;,
      \qquad i=1,\ldots,n\;,
\end{equation}
where $\kappa$ is a dimensional scaling constant for the detector, $n$ is 
the number of modulation states in the cycle, and $\delta\mathscr{I}_i$ 
is a random fluctuation due to photon noise statistics.

            It therefore makes sense to derive the expressions for the 
expectation value and variance of $\mathscr{I}_i$ -- respectively, 
$E(\mathscr{I}_i)$ and $\sigma^2(\mathscr{I}_i)$ -- as these provide 
important information about the quality of the observations 
\citep[see also][]{AC08}.
      As we anticipated earlier in this section, $E(\bm{x}(t))=0$. 
      Similarly, the photon noise fluctuation of the measurement has zero 
expectation value. 
      Therefore, using Eq.~(\ref{eq:signal_i}), and the fact that the 
modulation process is assumed to be fully deterministic and stationary, 
we obtain
\begin{eqnarray} \label{eq:av_signal_i}
\bar{\mathscr{I}}_i
&\equiv&E(\mathscr{I}_i)
	=\kappa\Delta t \sum_{j=1}^4 \Biggl[
      m_{ij}\,S_j +\frac{1}{\Delta t} 
      \int\limits_{-\Delta t/2}^{+\Delta t/2}
	m_j(t+t_i)\,\bm{\nabla} S_j\cdot E(\bm{x}(t+t_i))\;\textrm{d}t
      \Biggr] \nonumber \\
	&\equiv&\kappa\Delta t \sum_{j=1}^4 m_{ij}\,S_j\;,
	\qquad i=1,\ldots,n\;,
\end{eqnarray}
where we have defined
\begin{equation} \label{eq:mod_mat}
m_{ij}\equiv\frac{1}{\Delta t} 
      \int\limits_{-\Delta t/2}^{+\Delta t/2}m_j(t+t_i)\;\textrm{d}t\;.
\end{equation}
      The $n\times 4$ matrix $\tens{M}\equiv(m_{ij})$ so constructed is 
conventionally called the \textit{modulation matrix}.

      Equation~(\ref{eq:av_signal_i}) shows that the expectation value of 
the $i$-th modulated intensity signal is determined exclusively by the 
true Stokes vector $\bm{S}$, \textit{regardless of the modulation 
scheme adopted}.
      It is important to note that, in both Eqs.~(\ref{eq:av_signal_i}) 
and (\ref{eq:mod_mat}), the subscript $i$ no longer represents a specific 
instant $t_i$, like in Eq.~(\ref{eq:signal_i}), but strictly only the 
corresponding step position of the $n$-state modulation cycle.
      In particular, for Eq.~(\ref{eq:mod_mat}), this is a direct 
consequence of the stationarity of the modulation process.

      Similarly, the variance of $\mathscr{I}_i$, for $i=1,\ldots,n$,
is given by
\begin{eqnarray} \label{eq:sigma_modint}
\sigma^2(\mathscr{I}_i)
&\equiv& E([\mathscr{I}_i-\bar{\mathscr{I}}_i]^2)
	=E\Biggl(\Biggl[
      \kappa\sum_{j=1}^4
      \int\limits_{-\Delta t/2}^{+\Delta t/2}
	m_j(t+t_i)\,\bm{\nabla} S_j\cdot\bm{x}(t+t_i)\;\textrm{d}t
      +\delta\mathscr{I}_i\Biggr]^2\Biggr) \nonumber \\
&=&\kappa^2\sum_{j,k=1}^4\;\;
      \iint\limits_{-\Delta t/2}^{+\Delta t/2}
	m_j(t+t_i)\,m_k(t'+t_i)\,
      E([\bm{\nabla} S_j\cdot\bm{x}(t+t_i)]
      [\bm{\nabla} S_k\cdot\bm{x}(t'+t_i)])\;
      \textrm{d}t\,\textrm{d}t' \nonumber \\
&&+\sigma^2_\mathrm{p}(\mathscr{I}_i)\;,
\end{eqnarray}
where we indicated with $\sigma_\mathrm{p}(\mathscr{I}_i)$ the RMS
photon noise.

      Equation~(\ref{eq:sigma_modint}) has no immediate applicability to 
Stokes data analysis, since it relies on physical parameters, such as 
$\bm{\nabla}S_i$ and $\bm{x}(t)$, which are not directly available from 
typical observations.
      Instead, the variances defined by Eq.~(\ref{eq:sigma_modint}) 
are determined in practice from the statistics of repeated intensity 
measurements over many modulation cycles \citep{Li87}.
      Nonetheless, Eq.~(\ref{eq:sigma_modint}) helps clarifying the 
physical origin of the noise in the measurement of the modulated 
intensities, and therefore can provide insights on optimal choices of 
modulation schemes and frequencies for reducing the final error.

      We note that also in Eq.~(\ref{eq:sigma_modint}) the index $i$ no 
longer refers to a specific instant in time, $t_i$.
      Once again, this follows from the periodicity of $\bm{m}(t)$, and 
the stationarity of the random process described by $\bm{x}(t)$. 
      In fact, it is possible to drop altogether the temporal shift $t_i$ 
in the arguments of $\bm{x}(t)$ above, since the expectation value in the 
second line of Eq.~(\ref{eq:sigma_modint}) depends only on the difference 
of those two arguments, if $\bm{x}(t)$ is stationary (see 
Sect.~\ref{sec:covariances}).

      Following \cite{dTC00}, the optimal demodulation matrix (under
specific assumptions; see comments at the end of this section), which allows 
us to infer the incoming Stokes vector from the modulated intensity signals, 
is given by\footnote{This is in fact a general result that follows 
from error minimization in ordinary least squares problems
\citep[e.g.,][]{DS66}.}
$\tens{D}\equiv(d_{ij})=(\tens{M}^t\tens{M})^{-1}\tens{M}^t$.
      If we indicate the measured Stokes vector with $\bm{S}'$, we then have
\begin{equation} \label{eq:S_i}
S_i'=\sum_{j=1}^n d_{ij}\,\mathscr{I}_j\;,
	\qquad i=1,\ldots,4\;.
\end{equation}
      It is useful to derive the expectation value and variance also in 
the case of $S_i'$.
      Using Eq.~(\ref{eq:av_signal_i}), we find
\begin{eqnarray} \label{eq:av_S_i}
\bar S_i'
&\equiv& E(S_i')=\sum_{j=1}^n d_{ij}\,\bar{\mathscr{I}}_j
	=\kappa\Delta t \sum_{k=1}^4
	\Biggl(\sum_{j=1}^n d_{ij} m_{jk}\Biggr) S_k \nonumber \\
&=&\kappa\Delta t\,S_i\;,
	\qquad i=1,\ldots,4\;,
\end{eqnarray}
where we used the fact that $\sum_j d_{ij} m_{jk}=\delta_{ik}$.
      Similarly, for the variance of $S_i'$ we find, using 
Eqs.~(\ref{eq:S_i}) and (\ref{eq:av_S_i}),
\begin{eqnarray} \label{eq:var_S_i}
\sigma^2(S_i')
&\equiv& E([S_i'-\bar S_i']^2)
	=E\Biggl(\Biggl[\sum_{j=1}^n d_{ij}
	(\mathscr{I}_j-\bar{\mathscr{I}}_j)\Biggr]^2\Biggr) \nonumber \\
	&=&\sum_{j,k=1}^n d_{ij} d_{ik}\,
	E([\mathscr{I}_j-\bar{\mathscr{I}}_j]
	[\mathscr{I}_k-\bar{\mathscr{I}}_k])\;,
	\qquad i=1,\ldots,4\;.
\end{eqnarray}

      Equations (\ref{eq:av_signal_i})--(\ref{eq:sigma_modint}) and 
(\ref{eq:av_S_i}) and (\ref{eq:var_S_i}) provide the basic formulas through 
which we can evaluate the performance of different modulation schemes in 
the presence of atmospheric seeing.
      They will be related to earlier work in Sect.~\ref{sec:earlier}

      When the exposure time $\Delta t$ is sufficiently large compared to 
the coherence time of the atmospheric seeing, like in the case of slow 
modulation cycles, then we can assume that the covariance terms 
$E([\mathscr{I}_j-\bar{\mathscr{I}}_j][\mathscr{I}_k-\bar{\mathscr{I}}_k])$ 
in Eq.~(\ref{eq:var_S_i}) are negligible for $j\ne k$. 
      In such case the variance of $S_i'$ can be expressed directly as a 
diagonal quadratic form of the variances of the modulated intensity signals,
\begin{equation} \label{eq:var_S_diag}
\sigma^2(S_i')=\sum_{j=1}^n d_{ij}^2\,\sigma^2(\mathscr{I}_j)\;,
	\qquad i=1,\ldots,4\;.
\end{equation}
      In Sect.~\ref{sec:covariances} we provide a rigorous demonstration of 
the above statement.
      However, in many cases -- certainly when fast cameras with frames 
rates $\gtrsim 10\,\mathrm{Hz}$ are employed -- the time interval 
$t_{i+1}-t_i$ (whose inverse conventionally defines the modulation 
frequency) is of the same order or less than the time scale $t_0$ of the 
atmospheric seeing.
      As a result, the intensity variations induced by seeing in different 
states of the modulation cycle are in general statistically correlated, 
so one must use the more general expression of Eq.~(\ref{eq:var_S_i}).

      We note that Eq.~(\ref{eq:var_S_diag}) also holds in the absence 
of seeing, in which case $\sigma^2(\mathscr{I}_j)$ evidently reduces to 
just the contribution due to photon noise.
      Because the signals from different camera exposures are always 
statistically independent, no covariance terms arise in this case, and 
the photon noise only affects the diagonal terms of Eq.~(\ref{eq:var_S_i}).
      This provides a simple recipe to include photon noise in the results 
presented in this paper.
      Thus, for notational convenience, we will simply drop the 
photon-noise terms in all of the following treatment.

      Finally, we observe that the diagonality of $\sigma^2(S'_i)$, as 
expressed by Eq.~(\ref{eq:var_S_diag}), is a fundamental assumption in 
the derivation of the optimal form of demodulation matrices presented 
by \cite{dTC00}.
      In other words, the usual definition of the efficiency of polarization 
modulation schemes relies on the fact that seeing-induced noise be 
negligible with respect to photon noise, or at least that the 
seeing-induced covariances of the type appearing in Eq.~(\ref{eq:var_S_i}) 
be vanishing.
      Since that approach is based on the minimization of the noise of the 
Stokes measurements in the absence of systematic errors, which is expressed 
by a relation formally identical to Eq.~(\ref{eq:var_S_diag}), there
remains an unanswered question: how would the condition of optimality of 
the demodulation matrix change, if the more general expression of 
Eq.~(\ref{eq:var_S_i}) --  which takes into account the systematic errors 
due to the seeing -- were adopted. 
      This is a subject of research in its own right, which we are not 
going to address further in this paper.

\section{Dual-Beam Polarimetry} \label{sec:dualbeam}

      In the case of dual-beam polarimetry, one has two independent sets 
of $n$ measurements that can be combined into 2$n$ new intensity signals 
of the form $\mathscr{I}_i^\pm\equiv\mathscr{I}^a_i\pm\mathscr{I}^b_i$, 
where ``$a$'' and ``$b$'' refer to the two beams.
      For an ideal polarimeter, the modulation vector for the two beams 
satisfy the simple relations $m^a_1(t)=m^b_1(t)=1$, and $m^a_i(t)=-m^b_i(t)$ 
for $i=2,3,4$.
      However, in practical cases the two beams are never perfectly balanced.
      We take into account beam imbalance through the detector's gain factor 
in front of Eq.~(\ref{eq:signal_i}), and introduce new modulation vectors 
for the dual-beam system, $\bm{m}^+(t)$ and $\bm{m}^-(t)$, according to
\begin{equation} \label{eq:dual-beam matrix}
\bar\kappa\,\bm{m}^\pm_j(t)
      =\kappa_a \bm{m}^a(t)\pm\kappa_b \bm{m}^b(t)\;,
	\qquad \bar\kappa\equiv \kappa_a+\kappa_b\;.
\end{equation}
      Through these new vectors, 
Eqs.~(\ref{eq:av_signal_i})--(\ref{eq:sigma_modint}) can be directly 
extended to take into account the dual-beam redundancy.
      It is convenient to introduce a new $2n$ intensity vector with the 
corresponding modulation vector,
\begin{displaymath}
\mathscr{I}^\pm=(\mathscr{I}^+,\mathscr{I}^-)^T\;, \qquad 
      \bm{m}^\pm(t)=(\bm{m}^+(t),\bm{m}^-(t))^T\;.
\end{displaymath}
      We then find, for $i=1,\ldots,2n$,
\begin{eqnarray}
\label{eq:av_signal_db}
\bar{\mathscr{I}}^\pm_i
&\equiv& E(\mathscr{I}^\pm_i)
	=\bar\kappa\Delta t \sum_{j=1}^4 m^\pm_{ij}\,S_j\;, \\
\label{eq:var_signal_db}
\sigma^2(\mathscr{I}^\pm_i)
&\equiv& E([\mathscr{I}^\pm_i-\bar{\mathscr{I}}^\pm_i]^2) \\
&=&\kappa^2\sum_{j,k=1}^4\;\;
      \iint\limits_{-\Delta t/2}^{+\Delta t/2}
	m_j^\pm(t+t_i)\,m_k^\pm(t'+t_i)\,
      E([\bm{\nabla} S_j\cdot\bm{x}(t)]
      [\bm{\nabla} S_k\cdot\bm{x}(t')])\;\textrm{d}t\,\textrm{d}t'\;.
      \nonumber
\end{eqnarray}
      Note that we dropped the time shift $t_i$ from both arguments of 
$\bm{x}(t)$, relying on the stationarity of seeing (see the discussion 
after Eq.~[\ref{eq:sigma_modint}], and Sect.~\ref{sec:covariances}).
      The index of $t_i$ in the argument of the modulation functions 
$m_j^\pm(t)$ must be modulo $n$, since it is related to the actual time 
stamp during the modulation cycle.

      In the usual way, we can associate with 
$\tens{M}^\pm\equiv(m_{ij}^\pm)$ an ``optimal,'' dual-beam demodulation 
matrix, $\tens{D}^\pm$.
      Then, the extensions of Eqs.~(\ref{eq:av_S_i}) and (\ref{eq:var_S_i}) 
to the case of dual-beam polarimetry become, respectively,
\begin{eqnarray}
\label{eq:av_S_db}
\bar S_i'
&=&\sum_{j=1}^{2n}
	d^\pm_{ij} \bar{\mathscr{I}}_j^\pm\;,
	\qquad i=1,\ldots,4\;, \\
\label{eq:var_S_db}
\sigma^2(S_i')
&=&\sum_{j,k=1}^{2n} d^\pm_{ij}\,d^\pm_{ik}\,
	E([\mathscr{I}_j^\pm-\bar{\mathscr{I}}_j^\pm]
	[\mathscr{I}_k^\pm-\bar{\mathscr{I}}_k^\pm])\;,
	\qquad i=1,\ldots,4\;.
\end{eqnarray}
      As before, off-diagonal covariances in Eq.~(\ref{eq:var_S_db}) can 
be neglected only in the limit of exposure times much larger than the 
seeing coherence time ($\Delta t\gg t_0$).

      In order to illustrate the effect of beam imbalance, we consider 
the ideal modulation matrix for a general modulation scheme with $n$ states,
\begin{equation} 
\arraycolsep .3em
\tens{M}=%
\begin{pmatrix}
		1 &m_{12}&m_{13}&m_{14}\\
		\vdots &\vdots &\vdots &\vdots\\
		1 &m_{n2}&m_{n3}&m_{n4}\\
\end{pmatrix}\;.
\end{equation}
      If we introduce the degree of beam imbalance, 
$\rho=(\kappa_a-\kappa_b)/(\kappa_a+\kappa_b)$, the modulation matrix 
$\tens{M}^\pm$ appearing in Eq.~(\ref{eq:av_signal_db}) is accordingly 
given by
\begin{eqnarray} \label{eq:modulation_db}
\arraycolsep .3em
\tens{M}^\pm=%
\begin{pmatrix}
		1 &\rho\,m_{12}&\rho\,m_{13}&\rho\,m_{14}\\
		\vdots &\vdots &\vdots &\vdots\\
		1 &\rho\,m_{n2}&\rho\,m_{n3}&\rho\,m_{n4}\\
		\rho &m_{12}&m_{13}&m_{14}\\
		\vdots &\vdots &\vdots &\vdots\\
		\rho &m_{n2}&m_{n3}&m_{n4}\\
\end{pmatrix}\;.
\end{eqnarray}
where the top and bottom halves correspond respectively to the ``$+$'' 
and ``$-$'' linear combinations of the $n$ measurements from the two beams.

      From the form of $\tens{M}^\pm$ we see immediately that the seeing 
variations on the intensity and those on the polarization parameters of 
the incoming Stokes vector are decoupled when $\rho=0$.
      Such removal of cross-talk between intensity and polarization is 
in fact the rationale behind dual-beam polarimetry.
      However, this ideal goal is attained only if the intensity signals 
of the two beams can be balanced (either by camera gain adjustment, or by 
rescaling of the signals during data reduction) with an error which must 
be smaller than the target polarimetric sensitivity.
      To clarify this point, let us assume that the gain factors 
$\kappa_a$ and $\kappa_b$ have been experimentally determined with some 
error by the polarization calibration procedure, so that 
$\kappa_{a,b}=\kappa'_{a,b}+\delta \kappa'_{a,b}$, where $\kappa'_{a,b}$ 
is the measured value of $\kappa_{a,b}$.
      From the definition of the dual-beam modulation vector, 
Eq.~(\ref{eq:dual-beam matrix}), after rescaling, we have
\begin{displaymath}
\bar\kappa\,m^\pm_{ij}\equiv\frac{\kappa_a}{\kappa'_a}\,m^a_{ij}
      \pm\frac{\kappa_b}{\kappa'_b}\,m^b_{ij}\;,
\qquad \bar\kappa\equiv\frac{\kappa_a}{\kappa'_a}+\frac{\kappa_b}{\kappa'_b}\;,
\end{displaymath}
      Correspondingly the degree of beam imbalance becomes
\begin{displaymath}
\rho=\left(\frac{\delta \kappa'_a}{\kappa'_a}
      -\frac{\delta \kappa'_b}{\kappa'_b}\right)\biggl/
      \left(2+\frac{\delta \kappa'_a}{\kappa'_a}
      +\frac{\delta \kappa'_b}{\kappa'_b}\right)
\approx\frac{1}{2}\left(\frac{\delta \kappa'_a}{\kappa'_a}
      -\frac{\delta \kappa'_b}{\kappa'_b}\right)\;,
\end{displaymath}
so that $|\rho|\lesssim(1/2)(|\delta \kappa'_a|/\kappa'_a
      +|\delta \kappa'_b|/\kappa'_b)$ provides a sensible bound on the 
maximum error acceptable in order to achieve a given polarimetric precision 
in the presence of residual intensity-to-polarization cross-talk.

      Since the two beams can in principle always be rescaled 
\textit{post facto}, we will assume in the following that we are always 
dealing with perfectly balanced beams.
      It is understood that the possible rescaling of the two beams has to 
be taken into account for the proper determination of the noise on the the 
combined beams, according to the usual formula 
$\sigma^2(aX+bY)=a^2\sigma^2(X)+b^2\sigma^2(Y)$, where $a$ and $b$ are 
real numbers.
      This has an effect on the derivation of the quantities 
$\sigma^2(\mathscr{I}^\pm_j)$ in Eq.~(\ref{eq:var_signal_db}) from the 
noise statistics of the individual beams.
      In the appendix~\ref{sec:grating} we illustrate this problem for 
the particular case where the beam imbalance is produced by the 
differential efficiency of a diffraction grating in the $p$ and $s$ 
polarizations.

\section{The behavior of seeing-induced correlations between modulation states} \label{sec:covariances}

      We must evaluate covariance terms 
$E([\mathscr{I}_j-\bar{\mathscr{I}}_j][\mathscr{I}_k-\bar{\mathscr{I}}_k])$ 
in Eq.~(\ref{eq:var_S_i}) (or Eq.~[\ref{eq:var_S_db}]), and study their 
behavior as a function of the modulation frequency.
      Using the definition (\ref{eq:signal_i}),
\begin{eqnarray} \label{eq:first}
&&E([\mathscr{I}_j-\bar{\mathscr{I}}_j]
      [\mathscr{I}_k-\bar{\mathscr{I}}_k]) \\
&=&\kappa^2 \sum_{p,q=1}^4 
	E\left(\,
	\int\limits_{-\Delta t/2}^{+\Delta t/2}
	m_p(t+t_j)\,\bm{\nabla} S_p\cdot\bm{x}(t+t_j)\;\textrm{d}t
	\int\limits_{-\Delta t/2}^{+\Delta t/2}
	m_q(t'+t_k)\,\bm{\nabla} S_q\cdot\bm{x}(t'+t_k)\;\textrm{d}t'
	\right) \nonumber \\
&=&\kappa^2\Delta t^2
	\sum_{p,q=1}^4
	\partial_\alpha S_p\,\partial_\beta S_q\;
	\frac{1}{\Delta t^2}\;
	\iint\limits_{-\Delta t/2}^{+\Delta t/2}
	m_p(t+t_j)\,m_q(t'+t_k)\,
	E\bigl( x_\alpha(t+t_j)\,x_\beta(t'+t_k) \bigr)\;
      \textrm{d}t\,\textrm{d}t'\;, \nonumber
\end{eqnarray}
where in the last line a double summation over the coordinate indexes 
$\alpha,\beta=1,2$ is implicit.

      Because seeing can be considered a stationary random process, the 
expectation value in the last line of Eq.~(\ref{eq:first}) can be written 
in terms of the two-time correlation matrix,
\begin{equation} \label{eq:2time_corr}
\Gamma_{\alpha\beta}(t'-t)
	\equiv E\left(x_\alpha(t)\,x_\beta(t')\right)\;,
	\qquad \alpha,\beta=1,2\;,
\end{equation}
and because of the isotropy of the seeing motion, we also 
have\footnote{The cross-correlation function of a real stationary random 
process satisfies the relation 
$\Gamma_{\alpha\beta}(t)=\Gamma_{\beta\alpha}(-t)$.
The additional symmetry constraint provided by Eq.~(\ref{eq:symmetric}b) 
can be viewed as a consequence of the time-reversal symmetry of the 
seeing motion.}
\begin{subequations} \label{eq:symmetric}
\begin{eqnarray}
\Gamma_{11}(t)
&=&\Gamma_{22}(t)\equiv\Gamma(t)\;, \\
		\Gamma_{12}(t)
&=&\Gamma_{21}(t)\;,
\end{eqnarray}
\end{subequations}
since the two components of such a motion cannot be distinguishable.
      In addition, the two component motions $x_1(t)$ and $x_2(t)$ are 
orthogonal, and therefore they can be assumed to be independent random 
processes.
      Hence,
\begin{displaymath}
E(x_\alpha(t)x_\beta(t'))=E(x_\alpha(t)) E(x_\beta(t'))=0\;,
	\qquad \alpha\ne\beta\;,
\end{displaymath}
since $x_1(t)$ and $x_2(t)$ are random processes with zero average.
      Thus the correlation matrix is diagonal, and proportional to the 
unit matrix,\footnote{Another way to state this result is to consider 
that since the correlation matrix (\ref{eq:symmetric}) is symmetric, it 
can be diagonalized via a similarity transformation involving standard 
rotation matrices in O(2).
On the other hand, because of the isotropy of the seeing motion, there 
cannot be any preferential direction to attain such a diagonal form, 
and so the correlation matrix for the seeing displacement must always 
be diagonal.}
\begin{displaymath}
\Gamma_{\alpha\beta}(t)=\delta_{\alpha\beta}\,\Gamma(t)\;.
\end{displaymath}
      Equation~(\ref{eq:first}) then becomes
\begin{eqnarray} \label{eq:first1}
&&E([\mathscr{I}_j-\bar{\mathscr{I}}_j]
      [\mathscr{I}_k-\bar{\mathscr{I}}_k]) \\
&=&\kappa^2\Delta t^2
	\sum_{p,q=1}^4
	\bm{\nabla}S_p\cdot\bm{\nabla}S_q\;
	\frac{1}{\Delta t^2}\;
	\iint\limits_{-\Delta t/2}^{+\Delta t/2}
	m_p(t+t_j)\,m_q(t'+t_k)\,
	\Gamma(t'-t+t_k-t_j)\;\textrm{d}t\,\textrm{d}t'\;. \nonumber
\end{eqnarray}
      Following the formalism of Sect.~\ref{sec:dualbeam}, for a 
$n$-state modulation scheme in dual-beam configuration, $j$ and $k$ vary 
from 1 to $2n$.
      However, the indexes of $t_j$ and $t_k$ must be taken modulo $n$, 
because these refer to the actual steps of the modulation cycle.

      No further simplifications can be made at this point in the case of 
a continuously modulating device.
      In the remaining part of this section, we will then restrict 
ourselves to the case of stepped modulators, for which the following 
relation holds
\begin{displaymath}
m_i(t+t_j)=m_i(t_j)\equiv m_{ji}\;,
\qquad \forall t\in (-\Delta t/2,+\Delta t/2)\;.
\end{displaymath}
      The last equivalence follows directly from the definition of the 
modulation matrix, Eq.~(\ref{eq:mod_mat}), and so we can simply operate 
the substitution $m_p(t+t_j)\,m_q(t'+t_k)\to m_{jp}\,m_{kq}$ in 
Eq.~(\ref{eq:first1}).
      Using standard manipulations \citep[e.g.,][]{MW95}, the double 
integral in 
Eq.~(\ref{eq:first1}) can then be transformed into a single integral.
      In fact, noting that $t_k-t_j=(k-j)\Delta t/r$, where $r$ is the 
duty cycle of the camera ($0<r\le 1$), we have
\begin{eqnarray} \label{eq:single_int}
\mathscr{T}_s(\Delta t)
&\equiv&\frac{1}{\Delta t^2}\;
	\iint\limits_{-\Delta t/2}^{+\Delta t/2}
	\Gamma(t'-t+s\Delta t/r)\;\textrm{d}t\,\textrm{d}t' \nonumber \\
&=&\frac{1}{\Delta t}
	\int\limits_{-\Delta t}^{\Delta t}
	\left(1-\frac{|\tau|}{\Delta t}\right)
	\Gamma(\tau+s\Delta t/r)\;\textrm{d}\tau\;.
\end{eqnarray}

\begin{figure}[!t]
\centering
\includegraphics[width=5in]{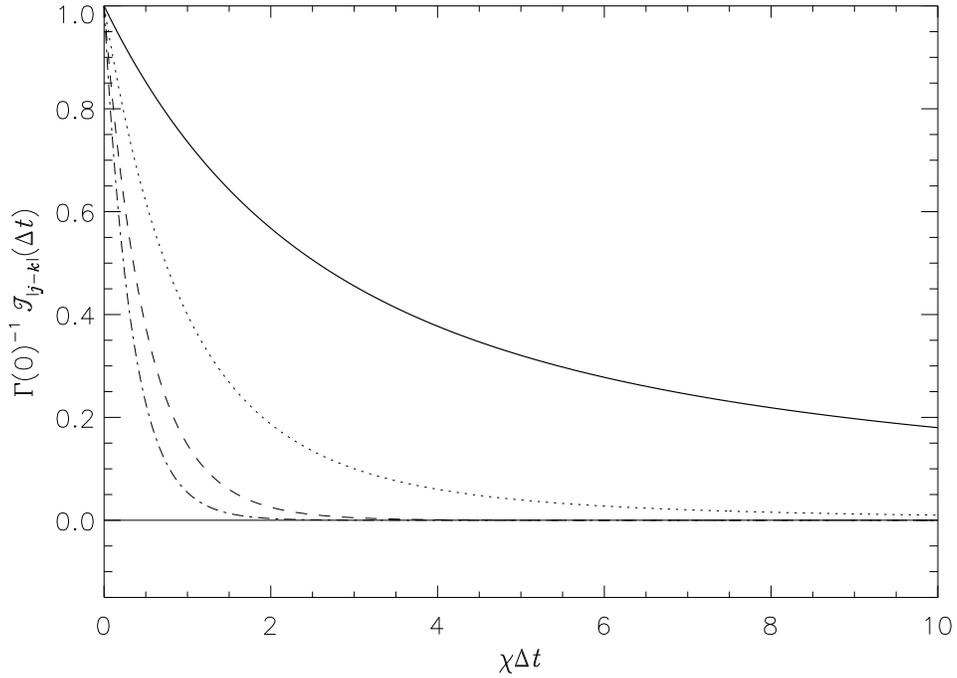}
\caption{Two-time correlations of the normalized 
amplitudes of the seeing displacement vector, 
$\mathscr{T}_{|j-k|}(\Delta t)$, plotted against the 
normalized exposure time, $\chi\Delta t$, for various values of $|j-k|$: 
0 (variance; thick continuous line); 1 (first neighbor; dotted line); 
2 (second neighbor; dashed line); 3 (third neighbor; dot-dashed line).
For this calculation, we used the expression (\ref{eq:Gamma}) for the 
auto-correlation function $\Gamma(t)$, and a duty cycle $r=1$ for the 
camera.}
\label{fig:correlations}
\end{figure}

      For the sake of demonstration, in the following we assume for 
$\Gamma(t)$ a functional dependence typical of a Gauss-Markov random 
process,
\begin{equation} \label{eq:Gamma}
\Gamma(t)=\Gamma(0)\,\mathrm{e}^{-\chi |t|}\;,\qquad\chi>0\;.
\end{equation}
      It is known \citep{Ta61} that the Kolmogorov description of 
atmospheric turbulence leads instead to an auto-correlation function of 
the form $\mathrm{e}^{-\chi|t|^{5/3}}$.
      However, its use in place of Eq.~(\ref{eq:Gamma}) would not change 
qualitatively the conclusions of this section, since Eq.~(\ref{eq:Gamma}) 
already contains the essential features of the seeing power spectrum that 
we are going to analyze.
      Using the auto-correlation function (\ref{eq:Gamma}), 
$\mathscr{T}_s(\Delta t)$ can be integrated analytically.
\begin{displaymath}
\mathscr{T}_s(\Delta t)=
\begin{cases} \displaystyle
	\Gamma(0)\,\frac{(\textrm{e}^{\chi\Delta t}-1)^2}{\chi^2\Delta t^2}\,
	\textrm{e}^{-\chi\Delta t\,(1+|s|/r)}\;,&\quad s\ne 0\;, \\
\noalign{\vspace{9pt}} \displaystyle
	\Gamma(0)\,\frac{2(\chi\Delta t+\textrm{e}^{-\chi\Delta t}-1)}%
	{\chi^2\Delta t^2}\;,&\quad s=0\;.
\end{cases}
\end{displaymath}
      We then see that $\mathscr{T}_s(\Delta t)\to 0$ when 
$\Delta t\gg\chi^{-1}$ as expected, because of the random nature of 
atmospheric seeing.
      In particular, for very long exposure times, 
$\mathscr{T}_s(\Delta t)$ tends to zero at least as 
$(\chi\Delta t)^{-2}$ for $s\ne 0$, whereas 
$\mathscr{T}_0(\Delta t)\sim (\chi\Delta t)^{-1}$.
      For typical atmospheric conditions, 
$\chi^{-1}\sim t_0\sim 0.01\,\mathrm{s}$, which is of the same order of 
magnitude of typical exposure times.
      \emph{Hence, the terms 
$E([\mathscr{I}_j-\bar{\mathscr{I}}_j][\mathscr{I}_k-\bar{\mathscr{I}}_k])$ 
must in general be taken into account for a proper determination of the 
measurement errors on the Stokes vector, even for $j\ne k$.}
      Figure~\ref{fig:correlations} shows how fast these off-diagonal 
terms drop as $|j-k|$ increases.
      A camera duty cycle with $r=1$ was assumed for that figure.

      In order to illustrate the effects of the covariances 
(\ref{eq:first1}) on the seeing noise on the measured Stokes parameters, 
we introduce the Stokes gradient matrix, 
$\tens{G}_{ij}=\bm{\nabla}S_i\cdot\bm{\nabla}S_j$, which allows us to 
recast Eq.~(\ref{eq:first1}) in matrix form,
\begin{equation} \label{eq:first2}
\tens{Cov}_{jk}(\Delta t)
\equiv E([\mathscr{I}_j-\bar{\mathscr{I}}_j]
	[\mathscr{I}_k-\bar{\mathscr{I}}_k])
	=\kappa^2\Delta t^2\,
	\left(\tens{M}\tens{G} \tens{M}^T\right)_{jk}
	\mathscr{T}_{|j-k|}(\Delta t)\;.
\end{equation}
      Here we used Eq.~(\ref{eq:single_int}), and the fact that 
$\mathscr{T}_{j-k}(\Delta t)=\mathscr{T}_{k-j}(\Delta t)$, as indicated by 
the explicit functional form of those integrals in practical cases.
      Consequently, Eq.~(\ref{eq:var_S_i}) can also be written in matrix 
form as
\begin{equation} \label{eq:var_S_matform}
\sigma^2(S_i')=\left(
	\tens{D}\,\tens{Cov}(\Delta t)\,\tens{D}^T
	\right)_{ii}\;.
\end{equation}

\begin{figure}[!t]
\centering
\includegraphics[width=5in]{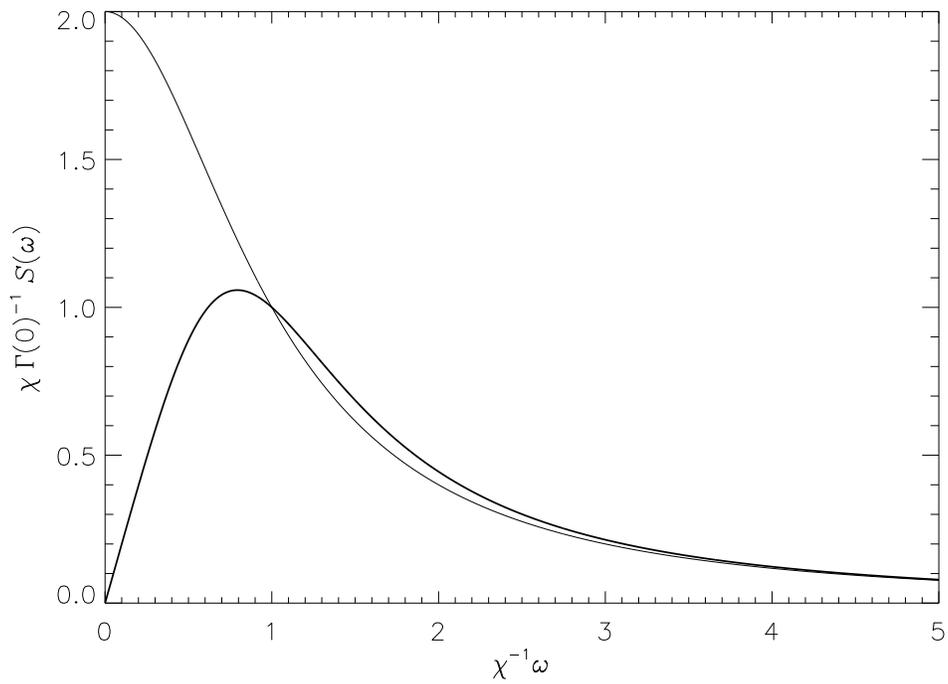}
\caption{Power spectra of a seeing-like random process.
The thin curve represents the normalized power spectrum corresponding to 
the auto-correlation function (\ref{eq:Gamma}).
The thick curve shows a simple analytic modification of this spectrum,
presenting characteristics similar to those produced by the action of
adaptive optics.}
\label{fig:spectrum}
\end{figure}

      We conclude this section by studying modifications of covariances 
(\ref{eq:first1}) arising from the use of adaptive optics.
      The general effect is an important reduction of the low-frequency 
part of the seeing power spectrum.
      Figure~\ref{fig:spectrum} shows an example based on the model of 
seeing correlations described by Eq.~(\ref{eq:Gamma}).
      Those curves should be compared qualitatively with the models 
of the seeing used by \cite{Ju04}.
      Here we define the power spectrum $S(\omega)$ as the Fourier
transform of the auto-correlation function \citep[e.g.,][]{MW95}
\begin{equation} \label{eq:power}
S(\omega)
	=\int\limits_{-\infty}^{+\infty}
	\Gamma(t)\,\textrm{e}^{-\mathrm{i}\,\omega t}\;\textrm{d}t
	=\Gamma(0)\int\limits_{-\infty}^{+\infty}
	\gamma(t)\,\textrm{e}^{-\mathrm{i}\,\omega t}\;\textrm{d}t\;,
\end{equation}
where $\gamma(t)=\Gamma(t)/\Gamma(0)$.
      The thin curve in Fig.~\ref{fig:spectrum} represents the power 
spectrum of the uncorrected seeing motion as described by the model of
Eq.~(\ref{eq:Gamma}), whereas the thick curve shows a simple analytic
modification of this spectrum with effects that could be representative 
of the action of adaptive optics.
      Note how the high-frequency part of the spectrum is not modified by 
the AO correction.

      In practical cases, the seeing will be described by an observed 
power spectrum, $S(\omega)$, rather than by a model auto-correlation 
function, $\Gamma(t)$.
      From Eq.~(\ref{eq:power}), we then have
\begin{equation} \label{eq:power_inv}
\Gamma(t)=\frac{1}{2\pi}\int\limits_{-\infty}^{+\infty}
	S(\omega)\,\mathrm{e}^{\mathrm{i}\,\omega t}\;\textrm{d}\omega\;,
\end{equation}
through which Eq.~(\ref{eq:single_int}) becomes
\begin{eqnarray} \label{eq:single_int_power}
\mathscr{T}_{j-k}(\Delta t)
&=&\frac{1}{2\pi\,\Delta t}
	\int\limits_{-\infty}^{+\infty} S(\omega)
	\int\limits_{-\Delta t}^{\Delta t}
	\left(1-\frac{|\tau|}{\Delta t}\right)
	\textrm{e}^{\mathrm{i}\,\omega[\tau+(k-j)\Delta t/r]}\;
      \textrm{d}\tau\,\textrm{d}\omega \nonumber \\
&=&\frac{1}{\pi}
	\int\limits_{-\infty}^{+\infty}S(\omega)\,
	\frac{1-\cos\omega\Delta t}{\omega^2\Delta t^2}\,
	\mathrm{e}^{\mathrm{i}\,\omega(k-j)\Delta t/r}\;\textrm{d}\omega\;.
\end{eqnarray}
      Because $S(\omega)=S(-\omega)$, it is easy to verify that 
Eq.~(\ref{eq:single_int_power}) implies 
$\mathscr{T}_{j-k}(\Delta t)=\mathscr{T}_{k-j}(\Delta t)$, as we had 
already derived earlier.

\begin{figure}[!t]
\centering
\includegraphics[width=5in]{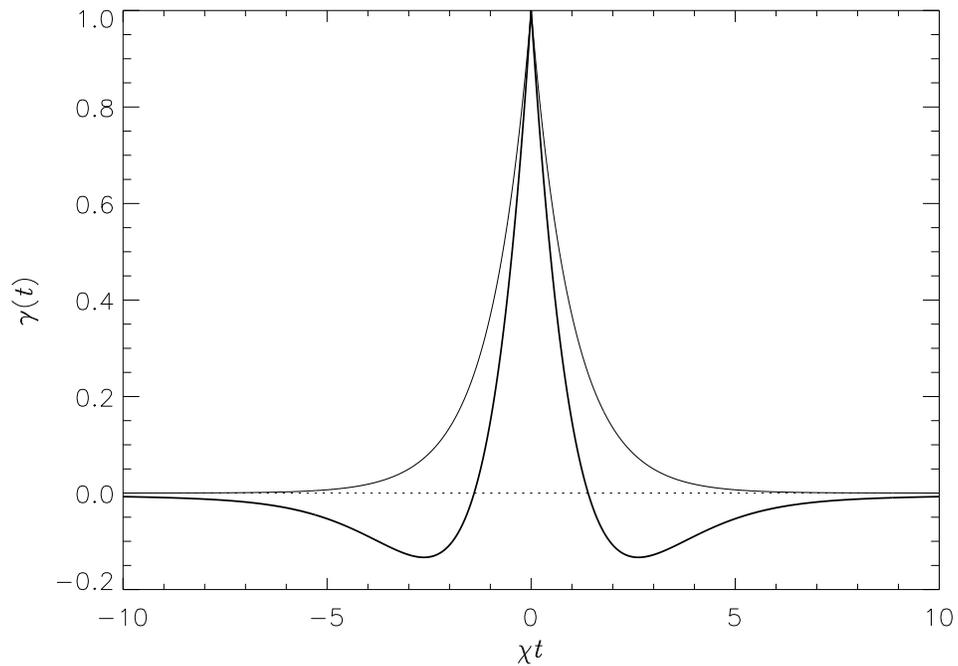}
\caption{Normalized auto-correlation function corresponding to the two 
power spectra of Fig.~\ref{fig:spectrum}.
The thin curve corresponds to Eq.~(\ref{eq:Gamma}) [after normalization 
by $\Gamma(0)$], whereas the thick curve corresponds to the 
AO-corrected spectrum.
Note the significant reduction of the time interval within which an 
efficient suppression of the auto-correlation of the seeing is attained 
in the presence of AO correction.}
\label{fig:autocorr}
\end{figure}

\begin{figure}[!t]
\centering
\includegraphics[width=5in]{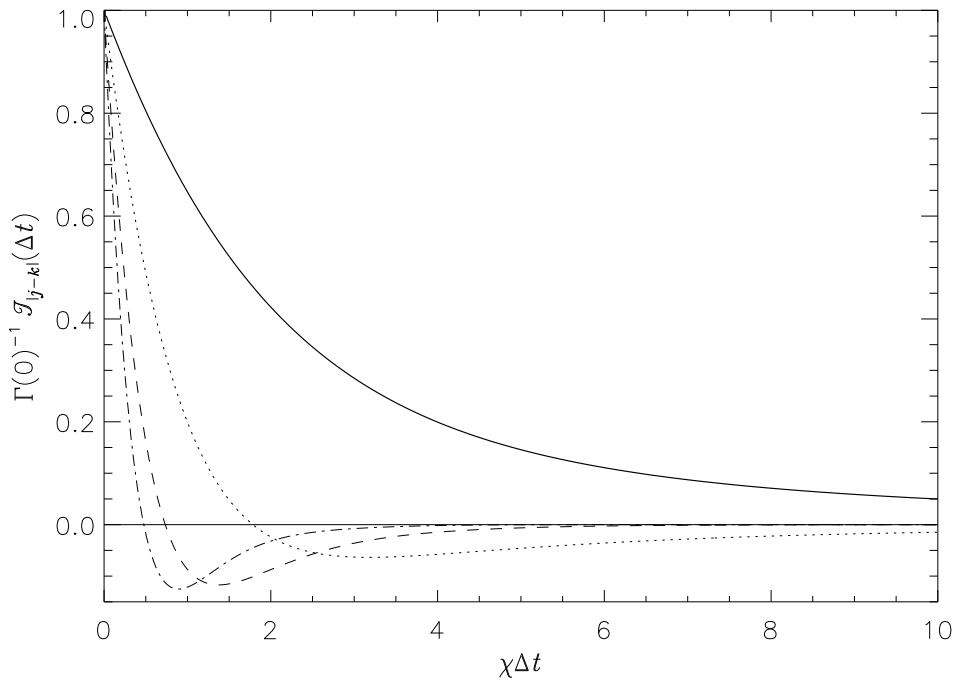}
\caption{Same as Fig.~\ref{fig:correlations}, but assuming the 
AO-corrected auto-correlation function shown in Fig.~\ref{fig:autocorr}.
Note the significant damping of the covariances for a given value of 
$\chi\Delta t$, compared to the case of uncorrected seeing. As for 
Fig.~\ref{fig:correlations}, a camera duty cycle $r=1$ was assumed for 
this calculation.}
\label{fig:correlations_AO}
\end{figure}

      The application of Eq.~(\ref{eq:power_inv}) to observed power 
spectra allows the determination of realistic auto-correlation functions 
of the seeing motion, even in the case of AO-corrected systems.
      As an example, Figure~\ref{fig:autocorr} shows the normalized 
auto-correlation functions corresponding to the two power spectra of 
Fig.~\ref{fig:spectrum}.
      Obviously, the thin curve corresponds to the auto-correlation 
function (\ref{eq:Gamma}).
      Once the true auto-correlation function for an AO-corrected system 
is known, one can use Eqs.~(\ref{eq:first1}) and (\ref{eq:single_int}) to 
determine how the modulated intensity covariances are modified by the AO 
correction.
      Figure~\ref{fig:correlations_AO} shows this effect in the case of 
the AO-corrected, auto-correlation function shown in Fig.~\ref{fig:autocorr} 
(thick curve).
      Comparing these results to those of Fig.~\ref{fig:correlations}, 
which correspond to the case of uncorrected seeing motion, we see that the 
variances drop faster in the presence of AO correction.
      On the other hand, the covariances do not converge to zero any 
faster than in the absence of AO correction.
      These covariances however change sign, so one could in principle 
adopt modulation schemes and frequencies such that some of the 
covariances are either vanishing, or even contributing a negative term 
to the expression of $\sigma^2(S_i')$, thus possibly reducing the final error.

\subsection{An illustrative example}
\label{sec:schemes}

      In order to illustrate the effect of seeing on polarization 
measurements, based on the results derived above, we will consider two 
popular, step-wise modulation schemes: a ``Stokes definition'' scheme and 
a ``balanced'' scheme, with ideal modulation matrices given respectively by
\begin{equation} \label{eq:schemes}
\arraycolsep .3em
\tens{M}_\mathrm{Sdef}\equiv
\begin{pmatrix}
		1 &+1 &0 &0 \\
		1 &-1 &0 &0 \\
		1 &0 &+1 &0 \\
		1 &0 &-1 &0 \\
		1 &0 &0 &+1 \\
		1 &0 &0 &-1
\end{pmatrix}\;,\quad
\tens{M}_\mathrm{bal}\equiv
\begin{pmatrix}
		1 &+\frac{1}{\sqrt3} &+\frac{1}{\sqrt3} &+\frac{1}{\sqrt3} \\
		1 &+\frac{1}{\sqrt3} &-\frac{1}{\sqrt3} &-\frac{1}{\sqrt3} \\
		1 &-\frac{1}{\sqrt3} &-\frac{1}{\sqrt3} &+\frac{1}{\sqrt3} \\
		1 &-\frac{1}{\sqrt3} &+\frac{1}{\sqrt3} &-\frac{1}{\sqrt3}
\end{pmatrix}\;.
\end{equation}
      Both schemes provide maximum modulation efficiencies 
for all Stokes parameters ($1$ for $S_1$, and  $1/\sqrt3$ for $S_i$, with 
$i\ge 2$).
      The respective modulation matrices in the dual-beam case are 
derived according to Eq.~(\ref{eq:modulation_db}).
      We consider here the case of perfect balancing of the two beams.

      In the case of the Stokes-definition scheme, by design, only one 
of the Stokes parameters contributes at any time to any given intensity 
signal combined from the two beams, so that all cross-talk terms between 
different Stokes parameters are eliminated.
      In addition, the seeing-induced Stokes variations enter the 
Stokes-definition scheme only through terms that are diagonal in 
$\bm{\nabla} S_j$.
      In the case of a balanced modulation scheme, instead, one must take 
into account general covariance terms that depend on both $\bm{\nabla} S_j$ 
and $\bm{\nabla} S_k$, for $j,k=2,3,4$, which complicate the expressions 
of the seeing-induced polarization cross-talk.

      For illustration consider a particular case with 
$\bm{\nabla}S_2=\bm{\nabla}S_3=0$, and $\bm{\nabla}S_1$ parallel to 
$\bm{\nabla}S_4$.
      This example might represent realistic distributions of magnetic 
fields on the solar surface, such as those of magnetic bright points 
associated with emerging flux, observed near disk center.
      Introducing then the quantities $g_1=|\bm{\nabla}S_1|$ and 
$g_4=|\bm{\nabla}S_4|$ in Eq.~(\ref{eq:var_S_matform}), we find, for the 
Stokes definition scheme,
\begin{subequations} \label{eq:stokes_def}
\begin{eqnarray}
\sigma^2(S_1')
&=&\textstyle{\frac{1}{6}}\,\kappa^2\Delta t^2\,g_1^2\left[
		\mathscr{T}_0(\Delta t)
		+\frac{1}{3}\sum_{s=1}^5 (6-s)\mathscr{T}_s(\Delta t)
		\right]\;, \\
		\sigma^2(S_2')
&=&\sigma^2(S_3')=0\;, \\
		\sigma^2(S_4')
&=&\textstyle{\frac{1}{2}}\,\kappa^2\Delta t^2\,g_4^2\left[
		\mathscr{T}_0(\Delta t)+\mathscr{T}_1(\Delta t)
		\right]\;,
\end{eqnarray}
\end{subequations}
while for the balanced modulation scheme,
\begin{subequations} \label{eq:balanced}
\begin{eqnarray}
\sigma^2(S_1')
&=&\textstyle{\frac{1}{4}}\,\kappa^2\Delta t^2\,g_1^2\left[
	\mathscr{T}_0(\Delta t)
	+\frac{1}{2}\sum_{s=1}^3 (4-s)\mathscr{T}_s(\Delta t)
	\right]\;, \\
\sigma^2(S_2')
&=&\textstyle{\frac{1}{4}}\,\kappa^2\Delta t^2\,g_4^2\left\{
	\mathscr{T}_0(\Delta t)-\mathscr{T}_2(\Delta t)
	-\frac{1}{2}[\mathscr{T}_1(\Delta t)-\mathscr{T}_3(\Delta t)]
	\right\}\;, \\
\sigma^2(S_3')
&=&\textstyle{\frac{1}{4}}\,\kappa^2\Delta t^2\,g_4^2\left\{
	\mathscr{T}_0(\Delta t)-\mathscr{T}_2(\Delta t)
	+\frac{1}{2}[\mathscr{T}_1(\Delta t)-\mathscr{T}_3(\Delta t)]
	\right\}\;, \\
\sigma^2(S_4')
&=&\textstyle{\frac{1}{4}}\,\kappa^2\Delta t^2\,g_4^2\left[
	\mathscr{T}_0(\Delta t)
	+\frac{1}{2}\sum_{s=1}^3 (4-s)\mathscr{T}_s(\Delta t)
	\right]\;,
\end{eqnarray}
\end{subequations}

      If we assume that the exposure time, $\Delta t$, is sufficiently 
large compared to the characteristic time of the seeing, $t_0$, then we 
can neglect all $\mathscr{T}_s(\Delta t)$ terms with $s\ne0$ in the 
above equations.
      In such case, we see that the two schemes are affected by the same 
\emph{total} error on the inferred Stokes vector, under identical conditions 
of camera exposure and time duration of the observation.
      This is because there are $6/4=1.5$ more modulation cycles for the 
balanced scheme than for the Stokes-definition scheme, during the same 
time interval.
      Equations~(\ref{eq:stokes_def}) also show that the seeing-induced 
error in the Stokes-definition scheme only affects the Stokes parameters 
that have non-vanishing gradients at the entrance of the modulator.
      In particular, $g_4$ only induces an error on $S_4$.
      In contrast, in the balanced scheme, the \emph{same} error is evenly 
distributed among all of $S_2$, $S_3$, and $S_4$. 

      It is important to remark that the variances expressed by 
Eqs.~(\ref{eq:stokes_def}) and (\ref{eq:balanced}) strictly apply to
Stokes vectors entering the modulator of a polarization-free telescope.
      In general, instead, typical telescopes' Mueller matrices map 
the gradient vector $(\bm{\nabla}S_1,0,0,\bm{\nabla}S_4)$ at the entrance
to the telescope onto a new gradient vector 
$(\bm{\nabla}S_1',\bm{\nabla}S_2',\bm{\nabla}S_3',\bm{\nabla}S_4')$ at 
the entrance to the modulator, so that all inferred Stokes parameters
are affected by seeing-induced errors.
      In such case, there is no evident advantage in adopting the 
Stokes-definition scheme over a balanced modulation scheme.

\section{The effect of seeing during a full observational sequence}
\label{sec:generalization}

      The results of Sect.~\ref{sec:covariances} apply to elemental 
observations that consist of a single modulation cycle.
      The typical mode of operation of slit-based spectro-polarimeters
\citep[and of some imaging polarimeters, such as the IMaX instrument 
on-board the Sunrise balloon mission;][]{MP11} is instead to integrate over 
multiple modulation cycles for each position of the slit in a map (or
wavelength position, in the case of imaging polarimeters).
      Even at moderate modulation frequencies, the seeing-induced 
correlations in adjacent modulation cycles do not vanish, so we cannot 
neglect covariance terms between different modulation states in different 
modulation cycles.
      Hence, we must regard the entire elemental observation as a single 
measurement.
      In order to do so, we consider the expression of the Stokes 
variances, Eq.~(\ref{eq:var_S_i}), and extend it to the case of a 
series $N$ of modulation cycles.
      Substituting Eq.~(\ref{eq:first1}), we find
\begin{eqnarray*}
\sigma^2(S_i')
&=&\sum_{j,k=1}^{nN} d_{ij} d_{ik}\,
	E([\mathscr{I}_j-\bar{\mathscr{I}}_j]
	[\mathscr{I}_k-\bar{\mathscr{I}}_k]) \\
&=&\kappa^2
	\sum_{p,q=1}^4
	\bm{\nabla}S_p\cdot\bm{\nabla}S_q
	\sum_{j,k=1}^{nN} d_{ij}d_{ik}\;
	\iint\limits_{-\Delta t/2}^{+\Delta t/2}
	m_p(t+t_j)\,m_q(t'+t_k)\,
	\Gamma(t'-t+t_k-t_j)\;\textrm{d}t\,\textrm{d}t'\;.
\end{eqnarray*}
      We then introduce the box function
\begin{displaymath}
\Pi_a(t)=
\begin{cases}
		1\;,& |t|\le a/2 \\
		0\;,& |t|> a/2
\end{cases}
\end{displaymath}
substitute Eq.~(\ref{eq:power_inv}), and finally operate the changes of 
variable $\xi=t+t_j$ and $\xi'=t'+t_k$.
      We find
\begin{eqnarray} \label{eq:sigma_tmp}
\sigma^2(S_i')
&=&\frac{\kappa^2}{2\pi}
	\sum_{p,q=1}^4
	\bm{\nabla}S_p\cdot\bm{\nabla}S_q
	\int\limits_{-\infty}^{+\infty} \textrm{d}\omega\,S(\omega)
	\int\limits_{-\infty}^{+\infty}
	\textrm{e}^{-\textrm{i}\omega\xi}
	\sum_{j=1}^{nN}
	d_{ij}\,\Pi_{\Delta t}(\xi-t_j)\,m_p(\xi)\;\textrm{d}\xi \nonumber \\
&&\kern 5cm\kern 1.78mm
      \times\int\limits_{-\infty}^{+\infty}
	\textrm{e}^{\textrm{i}\omega\xi'}
	\sum_{k=1}^{nN}
	d_{ik}\,\Pi_{\Delta t}(\xi'-t_k)\,m_q(\xi')\;\textrm{d}\xi'\;.
\end{eqnarray}

      To make this formula applicable to specific modulation schemes and 
seeing realizations, we must evaluate the Fourier transform 
$\tilde H_{ij}(\omega)$ of the functions
\begin{displaymath}
H_{ij}(\xi)=\frac{1}{\Delta t} \sum_{k=1}^{nN}
	d_{ik}\,\Pi_{\Delta t}(\xi-t_k)\,m_j(\xi)\;.
\end{displaymath}
      To this purpose, we first note that the elements $d_{ik}$ are 
periodic in $k$ with period $n$, so we can rewrite
\begin{eqnarray} \label{eq:H_ij}
H_{ij}(\xi)
&=&\frac{1}{\Delta t} \sum_{k=1}^n d_{ik} \sum_{l=0}^{N-1}
	\Pi_{\Delta t}(\xi-t_k-ln\Delta t/r)\,
	m_j(\xi) \nonumber \\
&=&\frac{1}{\Delta t} \sum_{k=1}^n d_{ik} \left(\sum_{l=0}^{N-1}
	\delta(\xi-t_k-ln\Delta t/r)\ast
	\bigl[\Pi_{\Delta t}(\xi)\,m_j(\xi+t_k)\bigr]\right)\;,
\end{eqnarray}
where in the last equivalence we also used the periodicity of $m_j(t)$.

      We take advantage of the convolution theorem of Fourier analysis to 
derive $\tilde H_{ij}(\omega)$.
      We note that $H_{ij}(\xi)$ is a real-valued function, so its 
Fourier transform has the conjugation property 
$\tilde H_{ij}^\ast(\omega)=\tilde H_{ij}(-\omega)$.
      Equation~(\ref{eq:sigma_tmp}) then becomes
\begin{eqnarray} \label{eq:sigma_fourier}
\sigma^2(S_i')
&=&\frac{\kappa^2\Delta t^2}{2\pi}
	\sum_{p,q=1}^4
	\bm{\nabla}S_p\cdot\bm{\nabla}S_q
	\int\limits_{-\infty}^{+\infty} S(\omega)\,
	\tilde H_{ip}(\omega)\tilde H_{iq}^\ast(\omega)\;\textrm{d}\omega 
      \nonumber \\
&=&\frac{\kappa^2\Delta t^2}{2\pi}
	\int\limits_{-\infty}^{+\infty} S(\omega)\,
	\bigl(
	\tens{\tilde H}(\omega)\tens{G}\tens{\tilde H}^\dagger(\omega)
	\bigr)_{ii}\;\textrm{d}\omega\;,
\end{eqnarray}
where $\tens{\tilde H}^\dagger(\omega)$ is the Hermitian conjugate 
of $\tens{\tilde H}(\omega)$.
      Using the symmetry properties of $\tens{\tilde H}(\omega)$ and 
$\tens{G}$, it is simple  to demonstrate that the integrand in 
Eq.~(\ref{eq:sigma_fourier}) is an even function of $\omega$.
      This allows us to restrict the integration domain to $[0,+\infty)$, 
which is where the observed power spectrum is naturally defined.

      The case of a continuously rotating modulator is summarized in the 
appendix~\ref{sec:rotating}, where we limit ourselves to providing the 
Fourier transforms of the functions $\Pi_{\Delta t}(\xi)\,m_j(\xi+t_k)$ 
that appear in Eq.~(\ref{eq:H_ij}).
      In the case of a stepped modulator, $m_j(\xi+t_k)=m_{kj}$.
      For the unit box and the ``windowed'' comb functions in 
Eq.~(\ref{eq:H_ij}), the following Fourier transform pairs hold, 
\begin{eqnarray*}
\Pi_a(\xi-\xi_0)
&\;\longleftrightarrow\;&
	a\,\textrm{e}^{-\textrm{i}\omega\xi_0}\,
	\textrm{sinc}(\omega a/2)\;, \\
	\sum_{l=0}^{N-1} \delta(\xi-\xi_0-la)
&\;\longleftrightarrow\;&
	N\,\textrm{e}^{-\textrm{i}\omega[\xi_0+(N-1)a/2]}\,
	\frac{\textrm{sinc}(N\omega a/2)}{\textrm{sinc}(\omega a/2)}\;.
\end{eqnarray*}
      We thus find, also noting that $t_k=t_1+(k-1)\Delta t/r$,
\begin{equation} \label{eq:H_ijFT}
	\tilde H_{ij}(\omega)=N\,
	\textrm{sinc}(\omega\Delta t/2)\,
      \frac{\textrm{sinc}(Nn\omega\Delta t/2r)}
	{\textrm{sinc}(n\omega\Delta t/2r)}\,
	\textrm{e}^{-\textrm{i}\omega[t_1+(N-1)n\Delta t/2r]}
	\sum_{k=1}^n d_{ik} m_{kj}\,
	\textrm{e}^{-\textrm{i}\omega(k-1)\Delta t/r}\;.
\end{equation}
      Note that $d_{ik}$ is an element of the demodulation matrix 
corresponding to the extended measurement of $N$ cycles.
      Such matrix contains a factor $1/N$ with respect to the analogous 
matrix for one cycle.
      Therefore, we can replace $d_{ik}$ with the standard (one-cycle) 
demodulation matrix, and drop the factor $N$ in front of Eq.~(\ref{eq:H_ijFT}).
      Because the matrix $\tens{\tilde H}(\omega)$ always appears in a 
product with $\tens{\tilde H}^\dagger(\omega)$, we can drop all common 
phase factors from its definition, so we can rewrite
\begin{equation} \label{eq:Hmat}
\tilde H_{ij}(\omega)
	=\textrm{sinc}(\omega\Delta t/2)
	\frac{\textrm{sinc}(\omega T/2)\,}
	{\textrm{sinc}(\omega T/2N)}
	\sum_{k=1}^n d_{ik} m_{kj}\,
	\textrm{e}^{-\textrm{i}(k-1)\omega T/Nn}\;,
\end{equation}
where we also used the fact that $Nn\Delta t/r=T$, i.e., the duration of 
one elemental observation.
      In the case of dual-beam polarimetry, the summation in
Eq.~(\ref{eq:Hmat}) is extended to $2n$, while at the same time $(k-1)$
in the exponential must be taken modulo $n$.

      We note that 
$\tilde H_{ip}(\omega)\tilde H_{iq}^\ast(\omega)\to\delta_{ip}\,\delta_{iq}$ 
for vanishing $\Delta t$.
      Therefore, only $\tens{G}_{ii}=|\bm{\nabla}S_i|^2$ contributes to 
$\sigma^2(S_i')$ in Eq.~(\ref{eq:sigma_fourier}) for $\Delta t\to0$.
      This is in complete agreement with the result that can be derived 
from Eqs.~(\ref{eq:first2}) and (\ref{eq:var_S_matform}) under the same 
limit conditions.
      The results presented by \cite{Li87} and \cite{Ju04} correspond 
evidently to the diagonal case $p=q$.

\begin{figure}[!t]
\centering
\includegraphics[width=5in]{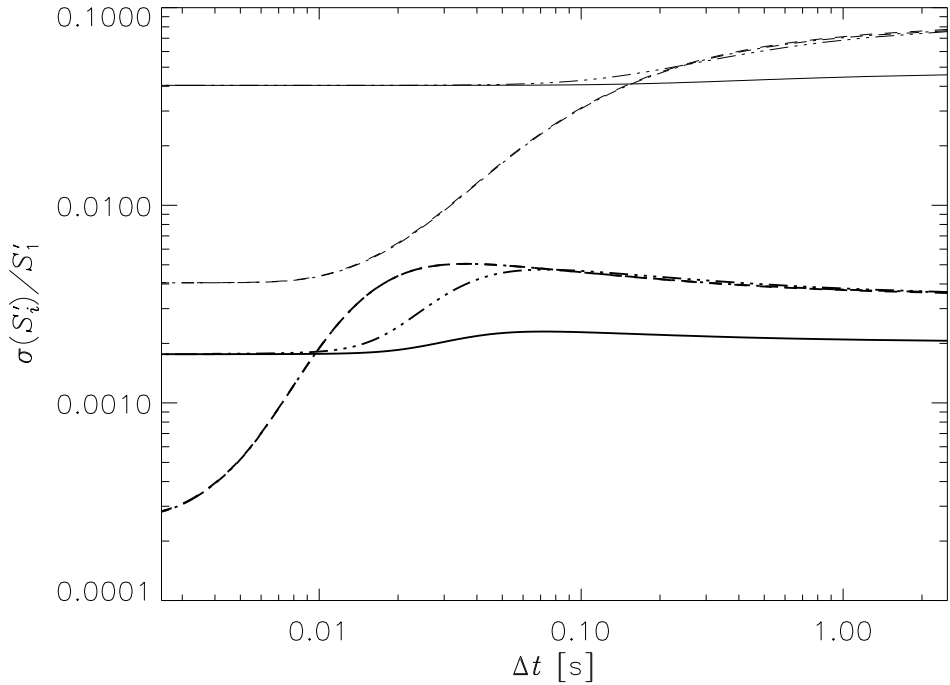}
\caption{Plots of the Stokes errors of Eq.~(\ref{eq:sigma_fourier}) 
normalized to the incoming intensity, as a function of the modulation
frequency.
The case shown corresponds to the balanced scheme of Eq.~(\ref{eq:schemes}) 
in single-beam configuration, with non-zero gradients of the Stokes 
parameters, such that $g_1=g_4=S_1\,\mathrm{arcsec}^{-1}$ and 
$g_2=g_3=0.1\,g_1$: \textit{continuous curve}, $\sigma(S_1')$; 
\textit{dashed curve}, $\sigma(S_2')$; \textit{dash-dotted curve}, 
$\sigma(S_3')$; \textit{dash-triple-dotted curve}, $\sigma(S_4')$.
These plots were calculated for a total modulation time $T=10\,\mathrm{s}$, 
assuming a camera duty cycle $r=1$.
The thin curves are for an observed power spectrum of the seeing, while 
the thick curves show the effects of AO correction (power spectra 
courtesy of T.~Rimmele).}
\label{fig:crosstalk.1beam}
\end{figure}

      Figure~\ref{fig:crosstalk.1beam} shows an example of 
seeing-induced errors on the measurement of the Stokes vector for a 
balanced modulation scheme in single-beam configuration, for a total 
modulation time of 10\,s, and with gradients 
$g_1=g_4=S_1\,\mathrm{arcsec}^{-1}$ and $g_2=g_3=0.1\,g_1$.
      Measured power spectra of seeing-induced image motions both with 
and without AO correction were used to produce these plots (T.~Rimmele, 
private communication).
      This figure shows the clear benefit of AO correction in terms of 
a reduction of the seeing-induced errors by more than an order of 
magnitude for $S_1$ and $S_4$.
      For modulation periods larger than the seeing correlation time, 
i.e., for long exposures, the variances of $S_2$, $S_3$, and $S_4$ are 
dominated by cross-talk from gradients in $S_1$.
      When the modulation frequency is increased to the point that 
seeing-induced displacements are practically frozen for the duration of 
a modulation cycle, the cross-talk terms in each error become negligible 
compared to the diagonal terms.
      The use of AO implies shorter correlation times of the 
seeing-induced displacements, and hence a higher modulation frequency is 
required to satisfy this condition.

\begin{figure}[!t]
\centering
\includegraphics[width=5in]{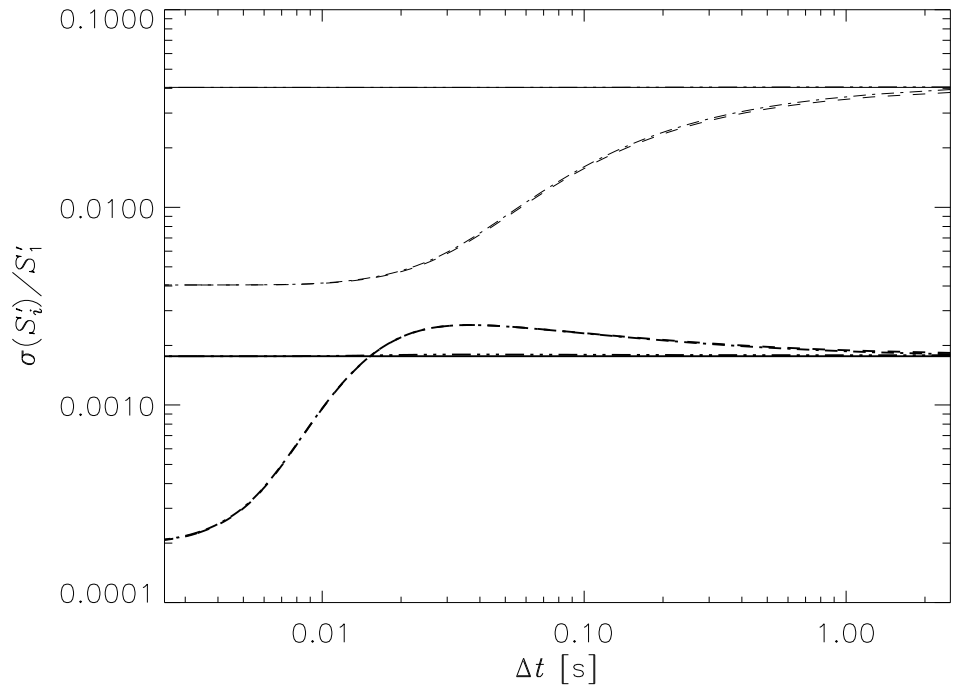}
\caption{Same as Fig.~\ref{fig:crosstalk.1beam}, but for the dual-beam 
configuration with perfectly balanced beams.}
\label{fig:crosstalk}
\end{figure}

      For the dual-beam case (Fig.~\ref{fig:crosstalk}) with perfectly 
balanced beams, there is no cross-talk from $S_1$ to $S_2$, $S_3$, and 
$S_4$ (see Sect.~\ref{sec:dualbeam}). This fact is illustrated by 
the drop in the error curves for $S_2$, $S_3$, and $S_4$ with respect to the
single-beam case of Fig.~\ref{fig:crosstalk.1beam}.
      However, the cross-talk among $S_2$, $S_3$, and $S_4$ remains.
      Analogously to the single-beam case, at high modulation frequencies, 
the seeing-induced error on each element of the Stokes vector tends to the 
contribution from the diagonal term only.
      Comparing the plots of Figures~\ref{fig:crosstalk.1beam} and 
\ref{fig:crosstalk}, we see that for very high modulation frequencies 
($\gtrsim100\,\mathrm{Hz}$ without AO, or $\gtrsim1\,\mathrm{kHz}$ with 
AO) the performance of single- and dual-beam modulation schemes become 
comparable, and the only benefit of a dual-beam polarimeter in such case 
is the redundancy of polarimetric information, leading to a reduction of 
the photon noise by a factor of $\sqrt{2}$ with respect to the 
single-beam case.

\begin{figure}[!t]
\centering
\includegraphics[width=5in]{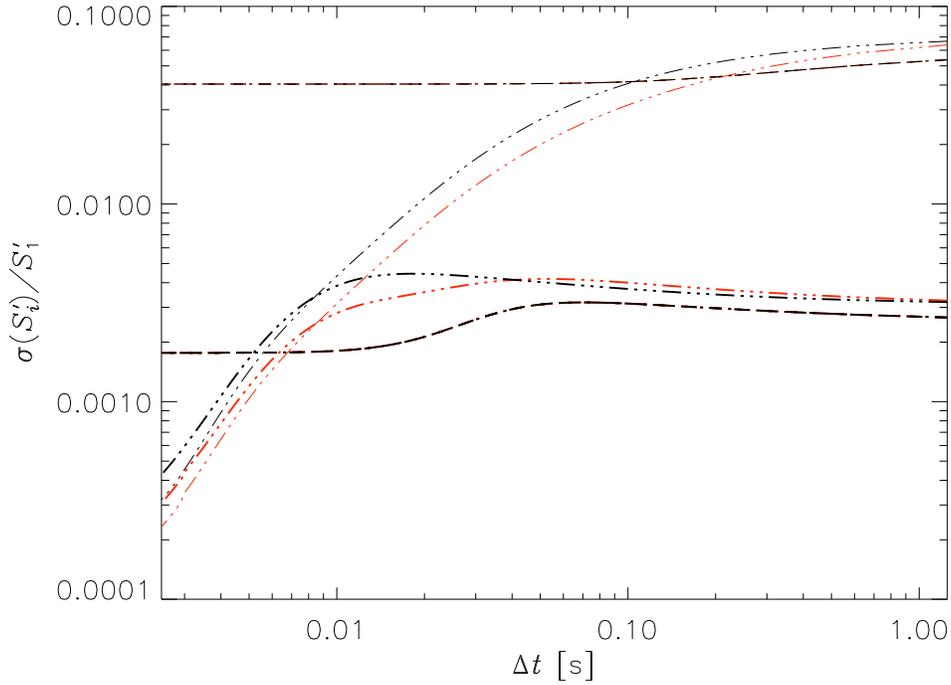}
\caption{Same as Fig.~\ref{fig:crosstalk}, but for the case of a 
stepped modulator (with 8 states) consisting of a waveplate with 
$150^\circ$ retardance, identical to the one considered in \cite{Li87}. 
The same demodulation scheme of \cite{Li87} was adopted for this 
calculation. These plots are for a dual-beam configuration with 
perfectly balanced beams. The gradients of the Stokes parameters are
$g_2=g_3=S_1\,\mathrm{arcsec}^{-1}$ and $g_1=g_4=0$. The red curves 
(see on-line version of the figure) show the predicted cross-talk level 
neglecting terms of the form $g_2\,g_3$, which are missing in the 
treatment of \cite{Li87,Ju04}.}
\label{fig:offdiag}
\end{figure}

      Our formalism produces results that agree qualitatively with 
those found by \cite{Li87} and \cite{Ju04}, although with some notable 
quantitative 
differences. Figure~\ref{fig:offdiag} shows the polarization cross-talk 
in a dual-beam configuration, calculated for a stepped modulator 
consisting of a rotating waveplate with $150^\circ$ retardance, 
identical to the one considered in the study by \cite{Li87}. The case 
considered in the figure corresponds to the presence of spatial 
gradients in $S_2$ and $S_3$ only. The red curves show the cross-talk 
that is derived by neglecting terms depending on 
$\bm{\nabla}S_2\cdot\bm{\nabla}S_3$, which are missing in the treatment 
by \cite{Li87} and \cite{Ju04} (see comment at the end of the next section).

\section{Discussion} \label{sec:earlier}

      In this paper, we have approached the determination of seeing-induced 
cross-talk noise from a statistical point of view. 
      Equations (\ref{eq:av_signal_i})--(\ref{eq:sigma_modint}) and 
(\ref{eq:av_S_i}) and (\ref{eq:var_S_i}) form the basis of our derivation 
(see Sect.~\ref{sec:formalism}), so it is important to compare those 
results with previous work \citep{Li87,Ju04}.

      In the work of \cite{Li87} and \cite{Ju04}, the ``variances'' 
there defined depend 
explicitly on the total observation time and the modulation frequency, 
through the integral of the product of the seeing power spectrum with a 
real function of frequency that depends implicitly on both (cf.~Eq.~[15] 
of \citealt{Li87}, and points 1--5 of Sect.~2 of \citealt{Ju04}).
      In our formalism, the total integration time does not appear 
explicitly because Eq.~(\ref{eq:sigma_modint}) represents the variance on 
a \textit{single} measurement of $\mathscr{I}_i$.
      However, if a number $N$ of such measurements are made, 
\textit{and those measurements can be considered statistically uncorrelated}, 
then the variance on the \textit{average} signal $\bar{\mathscr{I}}_i$ 
is reduced by a factor $N$.
      In such case, for a fixed modulation frequency, but increasing the 
total observation time, we expect the same qualitative scaling of variance 
with the integration time as derived in previous work.
      Secondly, the dependence on the modulation frequency and the seeing 
power spectrum is also apparent within our formalism -- already in the case 
of a single measurement -- when we consider Eq.~(\ref{eq:signal_i}).
      
      Incidentally, we note how the power spectrum of the seeing appears 
most naturally in our approach as the Fourier transform of the two-time 
correlation function of the seeing displacement vector (see 
Sect.~\ref{sec:covariances}), whereas in previous work it is identified 
instead with the modulus square of the Fourier transform of the seeing 
displacement.\footnote{We observe that the ordinary Fourier transform is 
not well defined for the seeing displacement vector, because the 
associated random process is not limited in time. 
One can get around this problem by defining a generalized Fourier 
transform of the seeing, as the limit of ordinary Fourier transforms of 
finite samples of the seeing for increasing duration of those samples 
\citep[e.g.,][]{MW95}.} 
      The correspondence between these two approaches to the definition 
of the power spectrum of a stationary random process is clearly 
described by \cite{MW95}.

      The importance of the covariances 
$E([\mathscr{I}_j-\bar{\mathscr{I}}_j][\mathscr{I}_k-\bar{\mathscr{I}}_k])$ 
in typical cases has fundamental implications for the concept of 
polarimetric measurements.
      Through our analysis we are able to quantify the significance of 
seeing-induced correlations between measurements corresponding to 
different modulation steps, and how these correlations decay for 
decreasing modulation frequencies.
      Based on those results, it must be expected that seeing-induced 
correlations between different modulation states typically extend 
beyond the time interval of one modulation cycle.
      In other words, the Stokes vector measurements corresponding to 
different modulation cycles during an elemental observation are in 
general statistically correlated.
      Under this condition, we must expect that the variance of the 
average signal $\bar{\mathscr{I}}_i$ will obey the ordinary scaling law 
by the total number $N$ of cycles of the elemental observation only 
approximately.
      Formally, one should consider instead such elemental observation 
as a \textit{single} measurement that is realized through the totality 
of the $N$ modulation cycles, and thus characterized by a corresponding 
$nN\times 4$ modulation matrix.
      The variance of this measurement can then be determined through 
the usual equations, adopting such extended definition of the modulation 
and demodulation matrices (see Sect.~\ref{sec:generalization}).
      A similar approach is taken also in the earlier work 
\citep{Li87,Ju04}, where these correlations are made manifest by performing 
a spectral analysis of the demodulated signal over the entire time 
interval of the elemental observation.

      Our formalism reveals however that, in the work of \cite{Li87} and
\cite{Ju04}, the 
covariance terms corresponding to contributions proportional to 
$\bm{\nabla}S_i\cdot\bm{\nabla}S_j$, with $i\ne j$ (see 
Eq.~[\ref{eq:first1}]), are not accounted for.
      In fact, \cite{Li87} derives the variances as Eq.~(12) 
of that paper directly from Eq.~(11), which are explicitly of diagonal form.
      In other words, looking at our Eqs.~(\ref{eq:first2}) and 
(\ref{eq:var_S_matform}), previous work has computed the seeing-induced 
noise always under the assumption that the gradient matrix $\tens{G}$ 
was diagonal.
      The additional off-diagonal components would arise in the development 
by \cite{Li87} and \cite{Ju04} when properly evaluating the expectation 
value of $(O_r - \bar O_r)^2$ in the notation of \cite{Li87}.
      Equations (\ref{eq:first2}) and (\ref{eq:var_S_matform}) also clarify 
the vanishing of the cross-talk terms in the limit of large modulation 
frequencies ($\chi\Delta t\to 0$).
      Because all the integrals $\mathscr{T}_{|j-k|}(\Delta t)$ tend to 
the same value in such limit, the vector of the Stokes variances becomes 
simply proportional to the diagonal of the gradient matrix, $\tens{G}$.

\section*{Acknowledgments}
We are grateful to our HAO colleague B.\ Lites for a careful reading of the
manuscript and helpful comments. We thank F.\ W\"oger (NSO) for
insightful discussion on the phenomenology of atmospheric seeing and of AO
corrections, and for a critical reading of Sect.~\ref{sec:seeing}.
We also acknowledge the valuable criticism and helpful comments by the
referees of a previous version of this manuscript.

\appendix

\section{Beam imbalance introduced by a polarizing grating}
\label{sec:grating}

      We consider a beam of unpolarized light incident on a diffraction 
grating.
      Typically the diffraction efficiency of a grating is different in 
the $p$ and $s$ directions, so the diffracted beam consists of a mixture 
of two orthogonally polarized beams with generally different intensities 
$S_{\rm p}$ and $S_{\rm s}$.
      Let us consider the case in which these two beams get separated 
by a perfectly polarizing beam-splitter placed after the grating.
      The emerging beams will then have intensity $S_{\rm p}$ and $S_{\rm s}$, 
respectively.

      Let us consider the combined signal of intensity $S$ so defined,
\begin{displaymath}
\bar\kappa S=\kappa_{\rm p} S_{\rm p} + \kappa_{\rm s} S_{\rm s}\;,
	\qquad \bar\kappa=\kappa_{\rm p}+\kappa_{\rm s}\;,
\end{displaymath}
where $\kappa_{\rm p}$ and $\kappa_{\rm s}$ are the two gain factors applied to the 
detector to balance the two beams.
      The ratio $\kappa_{\rm p}/\kappa_{\rm s}$ then corresponds to the ratio of the 
two orthogonal efficiencies of the grating, $\epsilon_{\rm p}$ and $\epsilon_{\rm s}$, 
according to
\begin{equation} \label{eq:app1}
\kappa_{\rm p}/\kappa_{\rm s}=\epsilon_{\rm s}/\epsilon_{\rm p}\;.
\end{equation}

      The photon noise associated with the combined signal $S$ is given by
\begin{eqnarray*}
\sigma^2(S)
&=&\left(\frac{\partial S}{\partial S_{\rm p}}\right)^{\!2}\sigma^2(S_{\rm p})
	+\left(\frac{\partial S}{\partial S_{\rm s}}\right)^{\!2}\sigma^2(S_{\rm s}) \\
&=&\left(\frac{\kappa_{\rm p}}{\bar\kappa}\right)^{\!2}\sigma^2(S_{\rm p})
	+\left(\frac{\kappa_{\rm s}}{\bar\kappa}\right)^{\!2}\sigma^2(S_{\rm s})\;.
\end{eqnarray*}
      Correspondingly, the relative error is given by
\begin{displaymath}
\frac{\sigma^2(S)}{S^2}
	=\left(\frac{\kappa_{\rm p} S_{\rm p}}{\bar\kappa S}\right)^{\!2}
	\frac{\sigma^2(S_{\rm p})}{S_{\rm p}^2}
	+\left(\frac{\kappa_{\rm s} S_{\rm s}}{\bar\kappa S}\right)^{\!2}
	\frac{\sigma^2(S_{\rm s})}{S_{\rm s}^2}\;.
\end{displaymath}
      Evidently, for balanced beams 
\begin{equation} \label{eq:app2}
\kappa_{\rm p} S_{\rm p}=\kappa_{\rm s} S_{\rm s}=\textstyle{\frac{1}{2}}\bar\kappa S\;,
\end{equation}
and therefore
\begin{equation} \label{eq:app3}
\frac{\sigma^2(S)}{S^2}
	=\frac{1}{4}\left[
	\frac{\sigma^2(S_{\rm p})}{S_{\rm p}^2}
	+\frac{\sigma^2(S_{\rm s})}{S_{\rm s}^2} \right]\;.
\end{equation}
      If we express the signals in terms of photon flux, assuming 
Poisson's statistics for the photon counts, and indicating with $r_N$ the 
read-out noise of the camera, we can rewrite Eq.~(\ref{eq:app3}) as
\begin{eqnarray} \label{eq:app4}
\frac{\sigma^2(S)}{S^2}
&=&\frac{1}{4}\left(
	\frac{N_{\rm p}+r_N^2}{N_{\rm p}^2}
	+\frac{N_{\rm s}+r_N^2}{N_{\rm s}^2} \right) \nonumber \\
&=&\frac{1}{4}\left(
	\frac{1}{N_{\rm p}}+\frac{1}{N_{\rm s}} \right)
	+\frac{r_N^2}{4}\left(
	\frac{1}{N_{\rm p}^2}
	+\frac{1}{N_{\rm s}^2} \right)\;.
\end{eqnarray}
      If we indicate with $N_\mathrm{in}$ the photon count before the 
grating, evidently $N_{\rm p}=\frac{1}{2}\epsilon_{\rm p} N_\mathrm{in}$ and 
$N_{\rm s}=\frac{1}{2}\epsilon_{\rm s} N_\mathrm{in}$, so we find
\begin{eqnarray} \label{eq:app5}
\frac{\sigma^2(S)}{S^2}
&=&\frac{1}{2 N_\mathrm{in}}\,\frac{\epsilon_{\rm p}+\epsilon_{\rm s}}{\epsilon_{\rm p}\epsilon_{\rm s}}
	+\frac{r_N^2}{N_\mathrm{in}^2}\,
	\frac{\epsilon_{\rm p}^2+\epsilon_{\rm s}^2}{\epsilon_{\rm p}^2\epsilon_{\rm s}^2} \nonumber \\
&=&\frac{1}{ N_\mathrm{in}}\,\frac{\bar\epsilon}{\epsilon_{\rm p}\epsilon_{\rm s}}
	+\frac{r_N^2}{N_\mathrm{in}^2}\,
	\frac{\epsilon_{\rm p}^2+\epsilon_{\rm s}^2}{\epsilon_{\rm p}^2\epsilon_{\rm s}^2}\;,
\end{eqnarray}
where we also introduced the grating's average efficiency 
$\bar\epsilon=\frac{1}{2}(\epsilon_{\rm p}+\epsilon_{\rm s})$.

      We note that the applicability of the above treatment in
practical cases relies on the previous knowledge of the grating
efficiencies, $\epsilon_{\rm p}$ and $\epsilon_{\rm s}$. These must be
determined as a function of wavelength from flat-fielding, ideally using 
unpolarized radiation. 
      If the incident beam is instead weakly polarized (as it is 
typically the case for sunlight flat-fields), the above treatment is 
still valid, but Eq.~(\ref{eq:app1}) can only provide an approximate 
estimate of the appropriate gain factors, $\kappa_{\rm s}$ and
$\kappa_{\rm p}$, if the measured efficiencies of the grating are 
adopted in that equation.

\section{Rotating modulator}
\label{sec:rotating}

      We consider the case of a retarding device, which is continuously 
rotating with angular frequency $\Omega$, and which at $0^\circ$ position 
is described by the Mueller matrix
\begin{displaymath}
\bm{\mu}_0\equiv
\begin{pmatrix}
      1 &0 &0 &0 \\
      0 &\mu_{22} &\mu_{23} &\mu_{24} \\
      0 &\mu_{32} &\mu_{33} &\mu_{34} \\
      0 &\mu_{42} &\mu_{43} &\mu_{44}
\end{pmatrix}\;,
\end{displaymath}
where the following norm conditions must be satisfied,
\begin{displaymath}
\sum_i\mu_{ij}^2\le 1\;,\qquad
\sum_j\mu_{ij}^2\le 1\;.
\end{displaymath}

      We note that a full modulation cycle corresponds to only half 
rotation of the modulator, because of the characteristic 
$180^\circ$-periodicity of polarization modulation.

      In order to determine the cross-talk terms (\ref{eq:sigma_fourier}) 
for such a device, we need to compute the Fourier transforms of the 
functions (see Eq.~[\ref{eq:H_ij}])
\begin{displaymath}
Z_{jk}(\xi)\equiv\Pi_{\Delta t}(\xi)\,m_j(\xi+t_k)\;,
      \qquad j=1,\dots,4\;,\quad k=1,\ldots,n\;, 
\end{displaymath}
where $n$ is the number of modulation states (camera exposures) in the 
modulation cycle.
      These Fourier transforms are given by the following expressions,
\begin{eqnarray*}
\tilde Z_{1k}(\omega)
&=&\Delta t\;\textrm{sinc}(\omega\Delta t/2)\;, \\
\tilde Z_{2k}(\omega)
&=&\frac{1}{2}\,(\mu_{22}+\mu_{33})\,\Delta t\;\textrm{sinc}(\omega\Delta t/2) \\
&&\kern .5cm
      +\frac{1}{4}\,[\mu_{22}-\mu_{33}+\mathrm{i}(\mu_{23}+\mu_{32})]\,
      \mathrm{e}^{4\mathrm{i}[\Omega t_1+\pi(k-1)/n]}\,
      \Delta t\;\textrm{sinc}\bigl([\omega-4\Omega]\Delta t/2\bigr) \\
&&\kern .5cm
      +\frac{1}{4}\,[\mu_{22}-\mu_{33}-\mathrm{i}(\mu_{23}+\mu_{32})]\,
      \mathrm{e}^{-4\mathrm{i}[\Omega t_1+\pi(k-1)/n]}\,
      \Delta t\;\textrm{sinc}\bigl([\omega+4\Omega]\Delta t/2\bigr)\;, \\
\tilde Z_{3k}(\omega)
&=&\frac{1}{2}\,(\mu_{23}-\mu_{32})\,\Delta t\;\textrm{sinc}(\omega\Delta t/2) \\
&&\kern .5cm
      -\frac{\mathrm{i}}{4}\,[\mu_{22}-\mu_{33}+\mathrm{i}(\mu_{23}+\mu_{32})]\,
      \mathrm{e}^{4\mathrm{i}[\Omega t_1+\pi(k-1)/n]}\,
      \Delta t\;\textrm{sinc}\bigl([\omega-4\Omega]\Delta t/2\bigr) \\
&&\kern .5cm
      +\frac{\mathrm{i}}{4}\,[\mu_{22}-\mu_{33}-\mathrm{i}(\mu_{23}+\mu_{32})]\,
      \mathrm{e}^{-4\mathrm{i}[\Omega t_1+\pi(k-1)/n]}\,
      \Delta t\;\textrm{sinc}\bigl([\omega+4\Omega]\Delta t/2\bigr)\;, \\
\tilde Z_{4k}(\omega)
&=&\frac{1}{2}\,(\mu_{24}+\mathrm{i}\mu_{34})
      \mathrm{e}^{2\mathrm{i}[\Omega t_1+\pi(k-1)/n]}\,
      \Delta t\;\textrm{sinc}\bigl([\omega-2\Omega]\Delta t/2\bigr) \\
&+&\frac{1}{2}\,(\mu_{24}-\mathrm{i}\mu_{34})
      \mathrm{e}^{-2\mathrm{i}[\Omega t_1+\pi(k-1)/n]}\,
      \Delta t\;\textrm{sinc}\bigl([\omega+2\Omega]\Delta t/2\bigr)\;,
\end{eqnarray*}
where $\Delta t$ is the exposure time for each modulation state, and $t_1$ 
is the time at which the modulator is found in the $0^\circ$ position.

\newpage

\end{document}